\documentclass{article}

\usepackage{custommacros}

\begin{document}




\title{Deterministic Objective Bayesian Analysis for Spatial Models}
\author{Ryan Burn \\ \href{mailto:ryan.burn@gmail.com}{ryan.burn@gmail.com}}
\date{}
\maketitle

\begin{abstract}
\cite{bergergp} and \cite{rengp} derived noninformative priors for Gaussian
process models of spatially correlated data using the reference prior
approach \citep{refprior}. The priors have good statistical properties and
provide a basis for objective Bayesian analysis \citep{objbay}. Using a
trust-region algorithm for optimization with exact equations for posterior
derivatives and an adaptive sparse grid at Chebyshev nodes, this paper develops
deterministic algorithms for fully Bayesian prediction and inference with the
priors. Implementations of the algorithms are available at
\href{https://github.com/rnburn/bbai}{https://github.com/rnburn/bbai}.
\end{abstract}

\section{Introduction}
Suppose we observe a Gaussian process $Z(\cdot)$ at sample points
$\s_1, \ldots, \s_n$ where
\begin{align}
	\Ex\cpa{Z(\s)} &= \beta_1 x_1(\s) + \cdots + \beta_p x_p(\s) \nonumber\\
		   &= \rg' \x(\s); \nonumber \\
	\Cov{Z(\s), Z(\u)} &= \var \cpa{\k(\norm{\s - \u}) + \nr};
  \label{eqn:gp}
\end{align}
$\x(\cdot)$ and $\k(\cdot)$ represent the known regressor function and correlation
function; and $\rg$, $\var$, $\l$, and $\nr$ represent the unknown regression coefficients,
signal variance, length, and noise-to-signal ratio.

Let $\params$ denote the unknown parameters $\pa{\rg, \var, \l, \nr}'$.
To reason about possible values at unobserved points, we'd like to know the distribution
$\pr{Z(\u)\gv \y, \params_\true}$
where $\u$ is an unobserved point and $\y$ denotes the observations
$\pa{Z(\s_1), \ldots, Z(\s_n)}'$.
Of course, different values of $\params$ could 
reasonably produce $\y$, so there's no way we can identify $\params_\true$ or construct prediction distributions 
exactly. We need ways to approximate.

\subsection*{Approach 1: Maximize Likelihood}
Suppose the likelihood function, $\like{\params} \propto \pr{\y\gv \params}$,
is strongly peaked about an optimum, $\params_\ml$. Then $\params_\true$ should
be close to $\params_\ml$, and $\pr{Z(\u)\gv\y, \params_\ml}$
should be a reasonable substitute for $\pr{Z(\u)\gv\y, \params_\true}$.

\begin{figure}
  \includegraphics{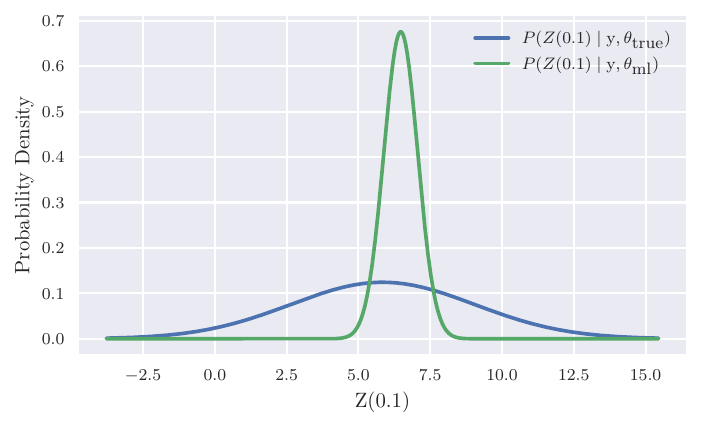}
  \caption{
  Compare Gaussian process prediction distributions $\pr{Z(0.1)\gv\y,\params_\ml}$ and 
$\pr{Z(0.1)\gv\y,\params_\true}$}
  \label{fig:ex1-ml-true}
\end{figure}

\begin{figure}
  \includegraphics{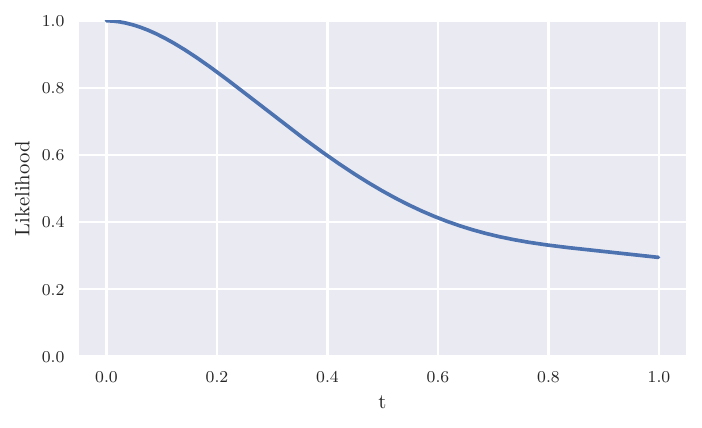}
  \caption{
Relative likelihood for different values of $\params$ on the line segment from $\params_\ml$ to $\params_\true$}
\label{fig:ex1-like}
\end{figure}

But what happens if a broad range of parameters could reasonably produce $\y$?
\begin{example}\label{ex:predictgpml}[\href{https://github.com/rnburn/bbai/blob/master/example/10-gaussian-process-bay-vs-ml.ipynb}{source}]
Consider the data set from Table~\ref{tab:ex1data}.
I randomly sampled the Gaussian process~\eqref{eqn:gp} with
\begin{align}
  \var = 25, \quad\l=0.01, \quad\nr=0.1,\;\;\textrm{and}\;\k(t)=\exp\cpa{-\frac{t^2}{2\l^2}}
  \label{eqn:ex1gp}
\end{align}
at 20 evenly spaced points on the interval $[0, 1]$. Likelihood has a maximum at
\begin{align*}
  \var_\ml = 34.42, \quad\l_\ml=0.035, \and\;\;\nr_\ml=3.82\times10^{-6}.
\end{align*}
Note how much smaller $\nr_\ml$ is than its true value. If we try to use
$\params_\ml$ as a substitute for $\params_\true$, we will get bad results as Figure~\ref{fig:ex1-ml-true} 
shows.
Put
\begin{align*}
	g(t) = L(\params_\ml (1 - t) + \params_\true t; \y) / L(\params_\ml; \y).
\end{align*}
$g(\cdot)$ computes the relative likelihood along a line segment from $\params_\ml$ to $\params_\true$, and Figure~\ref{fig:ex1-like} plots $g(t)$ for $0 \le t \le 1$. Looking at the figure, we can confirm that likelihood is not strongly peaked about an optimum and any value of $\params$ along the line segment could have reasonably produced
$\y$. 
\end{example}
\begin{table}
\centering
\begin{tabular}{lccccc}\toprule
  $i$ & $s$ & $y$ & $i$ & $s$ & $y$
	\\\cmidrule(lr){1-3} \cmidrule(lr){4-6}
  1 & 0.00 & 6.34 & 11 & 0.53 & 2.25 \\
  2 & 0.05 & 1.62 & 12 & 0.58 & 4.30 \\
  3 & 0.11 & 7.38 & 13 & 0.63 & -4.40 \\
  4 & 0.16 & 12.22 & 14 & 0.68 & -2.54 \\
  5 & 0.21 & 3.03 & 15 & 0.74 & 10.94 \\
  6 & 0.26 & -4.58 & 16 & 0.79 & -2.81 \\
  7 & 0.32 & -3.45 & 17 & 0.84 & -2.82 \\
  8 & 0.37 & -4.48 & 18 & 0.89 & 2.53 \\
  9 & 0.42 & -8.02 & 19 & 0.95 & 10.01 \\
  10 & 0.47 & 2.61 & 20 & 1.00 & 1.52
  \\\bottomrule
\end{tabular}
\caption{Randomly sampled data from Gaussian process~\eqref{eqn:ex1gp}}.
\label{tab:ex1data}
\end{table}

\subsection*{Approach 2: Integrate over Possible Parameters}
We saw in Example~\ref{ex:predictgpml} that using maximum likelihood parameters can lead
to poor results when the likelihood function isn't strongly peaked \citep{likeint}. Instead of approximating prediction distributions with only a single value of $\params$, let's
consider every $\params$ and weigh by a posterior distribution, $\post{\params}$,
\begin{align*}
	\prp{Z(\u)\gv\y} = \int \pr{Z(\u)\gv\y,\params} \post{\params}d\params.
\end{align*}
$\post{\params}$ measures our belief that parameters $\params$ generated the observations
$\y$. To derive $\post{\params}$, we apply Bayes' theorem:
$\post{\params} \propto \like{\params} \times \prior{\params}$
where $\prior{\params}$ measures our prior belief that the model has parameters $\params$.

Naturally, this leads to the question: How do we specify $\prior{\params}$ when
we know nothing particular about $\params$? Statisticians have grappled with the problem
of specifying so-called noninformative priors ever since Bayes and Laplace
first started applying the approach to the binomial model over 200 years ago.

While noninformative priors continue to be debated, fortunately, the
modern approach of reference priors gives a general path forward and,
particularly, for the case of Gaussian processes works quite well.

\begin{figure}
  \centering
  \includegraphics{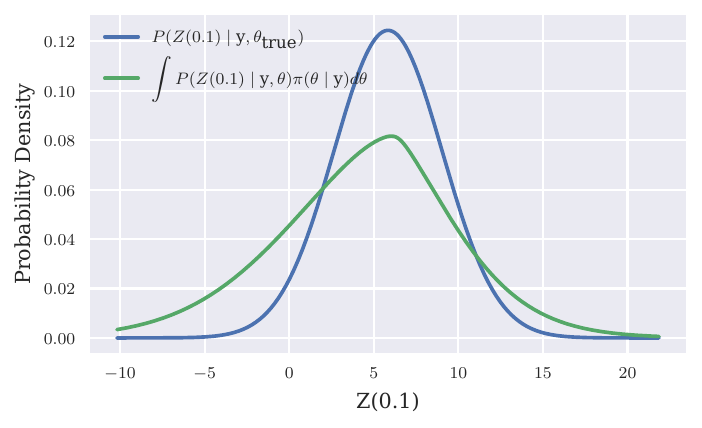}
  \caption{Compare the prediction distribution from the Bayesian approach 
    with reference prior, $\prp{Z(0.1)\gv\y}$,
  to the true prediction distribution $\pr{Z(0.1)\gv\y,\params_\true}$.}
\label{fig:ex1-bay-true}
\end{figure}

Before getting into the details (see \S\ref{sec:gpprior} and \S\ref{sec:detalgo} for descriptions of 
the prior and prediction algorithm), let's look at how the approach works on the Gaussian process
from Example~\ref{ex:predictgpml}.
\begin{example}\label{ex:predictgpbay} 
  [\href{https://github.com/rnburn/bbai/blob/master/example/10-gaussian-process-bay-vs-ml.ipynb}{source}]
  (Example~\ref{ex:predictgpml} continued)
In Figure~\ref{fig:ex1-bay-true}, I plot the prediction distribution for the 
Example~\ref{ex:predictgpml} data set using the Bayesian approach with a reference prior and compare to the
true prediction distribution.
We can see that the Bayesian approach gives a better approximation to the true prediction distribution than 
the maximum likelihood approach, Figure~\ref{fig:ex1-ml-true}.
\end{example}

\section{How to Specify Noninformative Priors}
The goal of a noninformative prior is to represent ``minimal information'' so that
inference is driven by the data and the model rather than prior knowledge.

Making this goal exact is difficult; and it's unlikely there will ever be a
universal approach to noninformative priors that's optimal for all situations, as there
can be multiple reasonable definitions of ``minimal information''. However, frequentist coverage
has emerged as one key metric to test whether a candidate noninformative prior is suitable for
objective Bayesian analysis. Here's the basic idea:
 Let $\pspace_1\times\cdots\times\pspace_k$ denote the parameter space for the model,
pick $\alpha$ to be something like $0.95$, and run Algorithm~\ref{alg:covfreq} for different $\tilde{\params}$ 
varied across the model's parameter space. If the prior is good, Algorithm~\ref{alg:covfreq} should
produce a result close to $\alpha$.
\begin{algorithm}[H]
  \caption{Test accuracy of credible sets produced with a prior \label{alg:covfreq}}
  \begin{algorithmic}[1]
    \Function{coverage-test}{$\tilde{\params}$, $j$, $\alpha$}
      \Let{$cnt$}{$0$}
      \Let{$N$}{a large number}
      \For{$i \gets 1 \textrm{ to } N$}
      	\Let{$\tilde{\y}$}{sample from $P(\cdot\gv \tilde{\params})$}
	\Let{$\tilde{\bs{\Theta}}$}{$
    \pspace_1\times\cdots\times\pspace_{j-1} 
    \times \pspace_j \cap (-\infty, \tilde{\theta_j}] \times
    \pspace_{j+1} \times \cdots \times \pspace_{k}
        $}
	\Let{$t$}{$\int_{\tilde{\bs{\Theta}}} \pi(\params\gv\tilde{\y}) d\params$}
	\If{$\frac{\alpha}{2} < t < 1 - \frac{\alpha}{2}$}
          \Let{$cnt$}{$cnt + 1$}
        \EndIf
      \EndFor
      \State \Return{$\frac{cnt}{N}$}
    \EndFunction
  \end{algorithmic}
\end{algorithm}

With Algorithm~\ref{alg:covfreq} in our toolbox, let's look at a few approaches for specifying
noninformative priors.

\subsection*{Constant Prior}
We begin with the simplest approach: Set $\prior{\params} \propto 1$.
Immediately, we see one serious disadvantage of this approach: It's not invariant under reparameterization. 
If $\varphi(\cdot)$ is some strictly increasing function onto $[a, b]$ with continuous derivative, then
\begin{align*}
	\int_a^b L(\theta; \y) d\theta = 
	\int_{\varphi^{-1}(a)}^{\varphi^{-1}(b)} L(\varphi(u); \y) \dot{\varphi}(u) du.
\end{align*}
Thus, different parameterizations with the constant prior lead to different posterior distributions. 

Still, let's try the approach out on some examples.
\begin{example}\label{ex:meancov-const}[\href{https://github.com/rnburn/bbai/blob/master/example/09-coverage-simulations.ipynb}{source}]
	Suppose we observe $n$ normally distributed values, $\y$, with variance $1$ and unknown mean, $\mu$.
	Then
	\begin{align*}
    L(\mu;\y) 
      &\propto \exp\cpa{-\frac{1}{2}\pa{\y - \mu \onev}'\pa{\y - \mu \onev}} \\
      &\propto \exp\cpa{-\frac{1}{2}\pa{n \mu^2 - 2 \mu n \bar{y}}} \\
      &\propto \exp\cpa{-\frac{n}{2}\pa{\mu - \bar{y}}^2}.
	\end{align*}
  Thus,
	\begin{align*}
    \int_{-\infty}^t \post{\mu} d\mu = 
    \frac{1}{2} \bra{1 + \erf\pa{\frac{t - \bar{y}}{\sqrt{2/n}}}}.
	\end{align*}
  I ran Algorithm~\ref{alg:covfreq} for $N=\num{10000}$, $\alpha=0.95$, and various values of
  $\mu$ and $n$. Table~\ref{tab:normal-mean-const-cov} shows the results.
\end{example}
\begin{table}
\centering
\begin{tabular}{lccccccl}\toprule
\multicolumn{2}{c}{$n = 5$} & 
\multicolumn{2}{c}{$n = 10$} &
\multicolumn{2}{c}{$n = 15$} &
\multicolumn{2}{c}{$n = 20$}
\\\cmidrule(lr){1-2}
\cmidrule(lr){3-4}
\cmidrule(lr){5-6}
\cmidrule(lr){7-8}
$\sigma^2$ & $\textrm{coverage}$ & 
$\sigma^2$ & $\textrm{coverage}$ &
$\sigma^2$ & $\textrm{coverage}$ &
$\sigma^2$ & $\textrm{coverage}$ 
\\\midrule
0.1 & 0.9502 & 0.1 & 0.9486 & 0.1 & 0.9508 & 0.1 & 0.9493 \\
0.5 & 0.9519 & 0.5 & 0.9478 & 0.5 & 0.9492 & 0.5 & 0.9488 \\
1.0 & 0.9516 & 1.0 & 0.9495 & 1.0 & 0.9517 & 1.0 & 0.9494 \\
2.0 & 0.9514 & 2.0 & 0.9521 & 2.0 & 0.9539 & 2.0 & 0.9489 \\
5.0 & 0.9489 & 5.0 & 0.9455 & 5.0 & 0.9558 & 5.0 & 0.9488
\\\bottomrule
\end{tabular}
\caption{Frequentist coverages for the mean of a normal distribution with known variance and constant prior.}
\label{tab:normal-mean-const-cov}
\end{table}

\begin{example}\label{ex:varcov-const}
  [\href{https://github.com/rnburn/bbai/blob/master/example/09-coverage-simulations.ipynb}{source}]
	Suppose we observe $n$ normally distributed values, $\y$, with zero-mean and unknown variance, $\sigma^2$.
	Then
	$L(\sigma^2;\y) \propto \pa{\frac{1}{\sigma^2}}^{n/2} \exp\cpa{-\frac{n s^2}{2 \sigma^2}}$
	where $s^2 = \frac{\y' \y}{n}$. Put $u = \frac{n s^2}{2 \sigma^2}$. Then
	\begin{align*}
		\int_0^t 
		\pa{\frac{1}{\sigma^2}}^{n/2} \exp\cpa{-\frac{n s^2}{2 \sigma^2}} d\sigma^2 
			&\propto \int_{\frac{n s^2}{2 t}}^\infty u^{n/2-2} \exp\cpa{-u} du \\
			&= \Gamma(\frac{n - 2}{2}, \frac{n s^2}{2 t}).
	\end{align*}
	Thus, 
	\begin{align*}
		\int_0^t \post{\sigma^2} d\sigma^2 = 
		  \frac{1}{\Gamma(\frac{n - 2}{2})}
			\Gamma(\frac{n - 2}{2}, \frac{n s^2}{2 t}).
	\end{align*}
  I ran Algorithm~\ref{alg:covfreq} for $N=\num{10000}$, $\alpha=0.95$, and various values of
  $\sigma^2$ and $n$. Table~\ref{tab:normal-var-const-cov} shows the results.
\end{example}
\begin{table}
  \centering
\begin{tabular}{lccccccl}\toprule
\multicolumn{2}{c}{$n = 5$} & 
\multicolumn{2}{c}{$n = 10$} &
\multicolumn{2}{c}{$n = 15$} &
\multicolumn{2}{c}{$n = 20$}
\\\cmidrule(lr){1-2}
\cmidrule(lr){3-4}
\cmidrule(lr){5-6}
\cmidrule(lr){7-8}
$\sigma^2$ & $\textrm{coverage}$ & 
$\sigma^2$ & $\textrm{coverage}$ &
$\sigma^2$ & $\textrm{coverage}$ &
$\sigma^2$ & $\textrm{coverage}$ 
\\\midrule
0.1 & 0.9014 & 0.1 & 0.9288 & 0.1 & 0.9418 & 0.1 & 0.9439 \\
0.5 & 0.9035 & 0.5 & 0.9309 & 0.5 & 0.9415 & 0.5 & 0.9398 \\
1.0 & 0.9048 & 1.0 & 0.9303 & 1.0 & 0.9404 & 1.0 & 0.9412 \\
2.0 & 0.9079 & 2.0 & 0.9331 & 2.0 & 0.9402 & 2.0 & 0.9393 \\
5.0 & 0.9023 & 5.0 & 0.9295 & 5.0 & 0.9339 & 5.0 & 0.9426
\\\bottomrule
\end{tabular}
\caption{Frequentist coverages for the variance of a normal distribution with known mean and constant prior.}
\label{tab:normal-var-const-cov}
\end{table}
In Example~\ref{ex:meancov-const}, the constant prior produces nearly perfect results. In 
Example~\ref{ex:varcov-const}, the prior is notably off for smaller values of $n$ but improves
as $n$ increases.

\subsection*{Jeffreys Prior}
Dissatisfied with the inconsistency of the constant prior under reparameterization, 
Harold Jeffreys searched for a better approach and proposed the prior
$\prior{\params} \propto \det{\I(\params)}^{1/2}$
where $\I(\params)$ is the Fisher information matrix,
\begin{align*}
  \I(\params)_{st} = \Ex_{\y}\cpa{
    \pa{\pd{\theta_s} \log \pr{\y\gv\params}}
    \pa{\pd{\theta_t} \log \pr{\y\gv\params}}
  }.
\end{align*}
We can check that unlike the constant prior, Jeffreys prior is invariant to reparameterization:
If $\bs{\varphi(\u)}$ is an injective continuously differentiable function whose range includes $\Theta$ and
whose Jacobian is never zero on $\bs{\varphi}^{-1}(\Theta)$, then the
change of variables formula gives us
\begin{align*}
  \int_{\Theta} \like{\params} \det{\I(\params)}^{1/2} d\params
    &= \int_{\bs{\varphi}^{-1}(\Theta)} \like{\bs{\varphi}(\u)} \det{\I(\bs{\varphi}(\u))}^{1/2} 
  \mid\pa{\det{\bs{D\varphi}(\u)}}\mid d\u,
\end{align*}
where $\bs{D\varphi}(\u)$ denotes the Jacobian matrix
$\bs{D\varphi}(\u)_{st} = \frac{\partial \varphi_s(\u)}{\partial u_t}$.
Let $\I^{\bs{\varphi}}(\u)$ denote the Fisher information matrix with respect to the 
reparameterization. Then
\begin{align*}
  \I^{\bs{\varphi}}(\u)_{st} 
    &= 
      \Ex_{\y}\cpa{
        \pa{\pd{u_s} \log\pr{\y\gv\bs{\varphi}(\u)}}
        \pa{\pd{u_t} \log\pr{\y\gv\bs{\varphi}(\u)}}
      } \\
    &= 
      \Ex_{\y}\cpa{
        \pa{\nabla_{\params}\log\pr{\y\gv\params}' \pdf{\bs{\varphi}}{u_s}(\u)}
        \pa{\nabla_{\params}\log\pr{\y\gv\params}' \pdf{\bs{\varphi}}{u_t}(\u)}
      } \\
    &=
    \pa{\pdf{\bs{\varphi}}{u_s}(\u)}' \I(\varphi(\u))
    \pa{\pdf{\bs{\varphi}}{u_t}(\u)}.
\end{align*}
Thus,
$\I^{\bs{\varphi}}(\u) = \bs{D\varphi}(\u)' \I(\bs{\varphi}(\u)) \bs{D\varphi}(\u)$
and
\begin{align*}
  \int_{\Theta} \like{\params} \det{\I(\params)}^{1/2} d\params
    &=
    \int_{\bs{\varphi}^{-1}(\Theta)} \like{\bs{\varphi}(\u)} \det{\I^{\bs{\varphi}}(\u))}^{1/2} d\u.
\end{align*}
\begin{example}\label{ex:meancov-jeff}(Example~\ref{ex:meancov-const} continued)
  To compute the Fisher information matrix, we first differentiate $\log\like{\mu}$,
  \begin{align*}
    \pd{\mu} \log \like{\mu} &= \pd{\mu}\pa{-\frac{n}{2}\pa{\mu - \bar{y}}^2} \\
                             &= -n \pa{\mu - \bar{y}}.
  \end{align*}
  Then we compute
  $\Ex_{\y}\cpa{\pa{\pd{\mu} \like{\mu}}^2\gv\mu} = 
    \Ex_{\y}\cpa{n^2 \pa{\mu - \bar{y}}^2\gv\mu}$.
  $\bar{y} - \mu$ is normally distributed with zero mean and variance $\inv{n}$, so
  $\Ex_{\y}\cpa{\pa{\pd{\mu} \like{\mu}}^2\gv\mu} = n$.
  Jeffreys prior in this case is the same as the constant prior.
\end{example}
\begin{example}\label{ex:varcov-jeff}
  [\href{https://github.com/rnburn/bbai/blob/master/example/09-coverage-simulations.ipynb}{source}]
  (Example~\ref{ex:varcov-const} continued)
  We differentiate $\log\like{\sigma^2}$ to get
  \begin{align*}
    \pd{\sigma^2} \log \like{\sigma^2} &= 
      \pd{\sigma^2}\pa{-\frac{n}{2}\log{\sigma^2} - \frac{n s^2}{2 \sigma^2}} \\
    &= \frac{n}{2\sigma^2}\pa{\frac{s^2}{\sigma^2} - 1}.
  \end{align*}
  Now,
  $\Ex_{\y}\cpa{\pa{\pd{\sigma^2} \like{\sigma^2}}^2\gv\sigma^2} = 
    \pa{\frac{n}{2\sigma^2}}^2
    \Ex_{\y}\cpa{\pa{\frac{s^2}{\sigma^2} - 1}^2\gv\sigma^2}$ and
   $y_1^2 + \cdots + y_n^2$
  follows a chi-squared distribution and with variance $2n\sigma^4$ and mean $n\sigma^2$, so
  \begin{align*}
    \Ex_{\y}\cpa{s^4\gv\sigma^2} &= \frac{\sigma^4\pa{2n + n^2}}{n^2} \\
                      &= \sigma^4\pa{1 + \frac{2}{n}}
  \end{align*}
  and
  \begin{align*}
    \Ex_{\y}\cpa{\pa{\pd{\sigma^2} \like{\sigma^2}}^2\gv\sigma^2} &= 
    \pa{\frac{n}{2\sigma^2}}^2
    \Ex_{\y}
    \cpa{\frac{s^4}{\sigma^4} - 2 \frac{s^2}{\sigma^2} + 1\gv\sigma^2} \\
    &= \pa{\frac{n}{2\sigma^2}}^2\pa{\frac{2}{n}} \\
    &= \frac{n}{2\sigma^4}.
  \end{align*}
  We derive the prior
  $\prior{\sigma^2} \propto \frac{1}{\sigma^2}$.
  For the CDF, we apply the same derivations in Example~\ref{ex:varcov-const} to get
  \begin{align*}
    \int_0^t \post{\sigma^2} d\sigma^2 = 
    \inv{\Gamma(\divtwo{n})}\Gamma(\divtwo{n}, \frac{n s^2}{2 t}).
  \end{align*}
  Using the same setup in Example~\ref{ex:varcov-const}, I produced the coverages in 
  Table~\ref{tab:normal-var-jeff-cov}.
\end{example}
\begin{table}
\centering
\begin{tabular}{lccccccl}\toprule
\multicolumn{2}{c}{$n = 5$} & 
\multicolumn{2}{c}{$n = 10$} &
\multicolumn{2}{c}{$n = 15$} &
\multicolumn{2}{c}{$n = 20$}
\\\cmidrule(lr){1-2}
\cmidrule(lr){3-4}
\cmidrule(lr){5-6}
\cmidrule(lr){7-8}
$\sigma^2$ & $\textrm{coverage}$ & 
$\sigma^2$ & $\textrm{coverage}$ &
$\sigma^2$ & $\textrm{coverage}$ &
$\sigma^2$ & $\textrm{coverage}$ 
\\\midrule
0.1 & 0.9516 & 0.1 & 0.9503 & 0.1 & 0.9509 & 0.1 & 0.9511 \\
0.5 & 0.9501 & 0.5 & 0.949 & 0.5 & 0.952 & 0.5 & 0.948 \\
1.0 & 0.9505 & 1.0 & 0.9511 & 1.0 & 0.9513 & 1.0 & 0.95 \\
2.0 & 0.948 & 2.0 & 0.9514 & 2.0 & 0.9501 & 2.0 & 0.9482 \\
5.0 & 0.9506 & 5.0 & 0.9497 & 5.0 & 0.9486 & 5.0 & 0.9485
\\\bottomrule
\end{tabular}
\caption{Frequentist coverages for the variance of a normal distribution with known mean and Jeffreys prior.}
\label{tab:normal-var-jeff-cov}
\end{table}
So far, Jeffreys prior performs excellently. In fact, for a single parameter, \cite{matchingprior}
show that in the limiting case, coverage for $\pa{1 - \alpha}\%$ credible sets using Jeffreys prior 
approaches $\alpha$ with an asymptotic error $o(n^{-1})$. Moreover, it's the only prior with this
property, so starting with the goal of matching coverage naturally leads us to Jeffreys prior.

Let's check how well Jeffeys prior performs in cases with more than a single variable.
\begin{example}\label{ex:meanvar-cov}
  [\href{https://github.com/rnburn/bbai/blob/master/example/09-coverage-simulations.ipynb}{source}]
	Suppose we observe $n$ normally distributed values, $\y$, with unknown mean, $\mu$, and unknown variance, $\sigma^2$.
	Then
	\begin{align*}
    L(\mu, \sigma^2;\y) \propto \pa{\frac{1}{\sigma^2}}^{n/2} \exp\cpa{
      -\inv{2\sigma^2}
      \pa{\y - \mu \onev}'
      \pa{\y - \mu \onev}
    }.
	\end{align*}
  We differentiate $\log \like{\cdot}$ to get
  \begin{align*}
    \pd{\mu} \log L(\mu, \sigma^2;\y)
      &= \frac{n}{\sigma^2} \pa{\bar{y} - \mu}\and \\
    \pd{\sigma^2} \log L(\mu, \sigma^2;\y)
      &= -\divtwo{n} \frac{1}{\sigma^2} + \half \pa{\inv{\sigma^2}}^2 
      \pa{\y - \mu \onev}'
      \pa{\y - \mu \onev}.
  \end{align*}
  We apply the derivations from Example~\ref{ex:meancov-jeff} and Example~\ref{ex:varcov-jeff} to
  get the Fisher information matrix
  $\I(\mu, \sigma^2) =
      \begin{pmatrix}
        \frac{n}{\sigma^2} & 0 \\
        0 & \frac{n}{2 \sigma^4} \\
      \end{pmatrix}
  $
  and the Jeffreys prior
  $\prior{\mu, \sigma^2} \propto \pa{\inv{\sigma^2}}^{3/2}$.
  Let's check coverage for $\sigma^2$. First, we integrate out $\mu$,
  \begin{align*}
    \int_{-\infty}^\infty \like{\mu, \sigma^2} &\prior{\mu, \sigma^2} d\mu \\
      &\propto
    \int_{-\infty}^\infty 
    \pa{\inv{\sigma^2}}^{(n+3)/2} \exp\cpa{-\inv{2 \sigma^2} \norm{\y - \mu \onev}^2 }d\mu \\
      &\!\begin{multlined}
      =
        \pa{\inv{\sigma^2}}^{(n+3)/2} 
    \exp\cpa{-\inv{2 \sigma^2} \pa{\y'\y - n \bar{y}^2}} \\
    \int_{-\infty}^\infty 
    \exp\cpa{-\frac{n}{2 \sigma^2} \pa{\mu - \bar{y}}^2}d\mu \\
    \end{multlined} \\
      &\propto
      \pa{\inv{\sigma^2}}^{(n+2)/2} 
    \exp\cpa{-\inv{2 \sigma^2} \pa{\y'\y - n \bar{y}^2}}.
  \end{align*}
  Then
  \begin{align*}
    \int_0^t \int_{-\infty}^\infty \post{\mu, \sigma^2} d \mu d\sigma^2 = 
    \inv{\Gamma(\divtwo{n})}\Gamma(\divtwo{n}, \inv{2 t}\pa{\y' \y - n \bar{y}^2}).
  \end{align*}
  I ran Algorithm~\ref{alg:covfreq} for $N=\num{10000}$, $\alpha=0.95$, $\mu=0$, and various values 
  of $\sigma^2$ to get the results in Table~\ref{tab:normal-var-jeff-cov2}.
\end{example}
\begin{table}
\centering
\begin{tabular}{lccccccl}\toprule
\multicolumn{2}{c}{$n = 5$} & 
\multicolumn{2}{c}{$n = 10$} &
\multicolumn{2}{c}{$n = 15$} &
\multicolumn{2}{c}{$n = 20$}
\\\cmidrule(lr){1-2}
\cmidrule(lr){3-4}
\cmidrule(lr){5-6}
\cmidrule(lr){7-8}
$\sigma^2$ & $\textrm{coverage}$ & 
$\sigma^2$ & $\textrm{coverage}$ &
$\sigma^2$ & $\textrm{coverage}$ &
$\sigma^2$ & $\textrm{coverage}$ 
\\\midrule
0.1 & 0.9241 & 0.1 & 0.938 & 0.1 & 0.9419 & 0.1 & 0.9463 \\
0.5 & 0.9219 & 0.5 & 0.9377 & 0.5 & 0.946 & 0.5 & 0.9441 \\
1.0 & 0.9245 & 1.0 & 0.94 & 1.0 & 0.9431 & 1.0 & 0.944 \\
2.0 & 0.9236 & 2.0 & 0.9391 & 2.0 & 0.9446 & 2.0 & 0.9432 \\
5.0 & 0.9182 & 5.0 & 0.9395 & 5.0 & 0.9403 & 5.0 & 0.9458
\\\bottomrule
\end{tabular}
\caption{Frequentist coverages for the variance of a normal distribution with unknown mean and Jeffreys prior.}
\label{tab:normal-var-jeff-cov2}
\end{table}
Unfortunately, the multiparameter case is not so easy; and as we see in Example~\ref{ex:meanvar-cov},
Jeffreys prior doesn't perform nearly as well. Jeffreys considered modifications of his prior
to handle the multiparameter case better but never developed a rigorous approach. For that, we turn
to reference priors.

\subsection*{Reference Priors}
If Jeffreys prior works well in the single parameter case, why not apply it to
parameters one at a time? In the reference prior approach \citep{refprior}, we build up a
multiparameter prior by marginalizing the likelihood with a conditional prior
of fewer parameters to form a new integrated likelihood function with only a single
parameter, to which we can apply Jeffreys prior.

Suppose $\like{\theta_1, \theta_2}$ is a likelihood function of two variables. We fix $\theta_1$
and use Jeffreys' approach to derive a conditional prior $\prior{\theta_2\gv\theta_1}$. Then we
integrate out $\theta_2$,
\begin{align*}
  \likei{\theta_1} = \int_{\Theta_2} \like{\theta_1, \theta_2} \prior{\theta_2\gv\theta_1} d\theta_2,
\end{align*}
to get the integrated likelihood function $\likei{\cdot}$ of only a single variable. We apply Jefferys'
approach again to the integrated likelihood function to get $\prior{\theta_1}$ and form the complete
prior
$\prior{\theta_1, \theta_2} = \prior{\theta_1} \times \prior{\theta_2\gv\theta_1}$.
If the prior $\prior{\cdot\gv\theta_1}$ is improper, we can choose a sequence of compact subsets
$A_1 \subset A_2 \subset \cdots \subset \Theta_2$
such that
$\lim_{t\to\infty} A_t = \Theta_2$,
apply the approach to $A_t$, and take the limit as $t\to\infty$.

Let's try this out on Example~\ref{ex:meanvar-cov}.
\begin{example}\label{ex:meanvarref-cov}
  [\href{https://github.com/rnburn/bbai/blob/master/example/09-coverage-simulations.ipynb}{source}]
  (Example~\ref{ex:meanvar-cov} continued). We first integrate out $\mu$ using the constant
  conditional prior,
  \begin{align*}
    \likei{\sigma^2} &= \int_{-\infty}^\infty \like{\mu, \sigma^2} \prior{\mu\gv\sigma^2} d\mu \\
      &\propto
      \pa{\inv{\sigma^2}}^{n/2} 
      \exp\cpa{-\inv{2\sigma^2}\pa{\y'\y - n\bar{y}^2}}
      \int_{-\infty}^\infty
        \exp\cpa{-\frac{n}{\sigma^2} \pa{\mu - \bar{y}}^2}
        d\mu \\
                          &\propto 
      \pa{\inv{\sigma^2}}^{(n-1)/2} 
      \exp\cpa{-\inv{2\sigma^2}\pa{\y'\y - n\bar{y}^2}}.
  \end{align*}
  Now, we differentiate $\likei{\cdot}$ to find the Fisher information matrix,
  \begin{align*}
    \pd{\sigma^2} \log \likei{\sigma^2}
      &= -\frac{n-1}{2\sigma^2} + \inv{2}\pa{\inv{\sigma^2}}^2 
        \pa{\y'\y - n\bar{y}^2} \\
      &= \inv{2\sigma^2}\cpa{\inv{\sigma^2}\pa{\y'\y - n\bar{y}^2} - (n-1)}.
  \end{align*}
  Put $Z = \frac{1}{\sigma^2}\pa{\y'\y - n\bar{y}^2}$.
  Then $Z$ follows a chi-squared distribution with $n-1$ degrees of freedom so that
    $\Ex[Z] = n-1$,
    $\Ex[Z^2] = 2 (n - 1) + (n - 1)^2$,
  and
  \begin{align*}
    \I(\sigma^2) &= \pa{\inv{2\sigma^2}}^2\cpa{\Ex[Z^2] -2(n-1)\Ex[Z] + (n-1)^2} \\
                 &= \frac{n - 1}{2 \sigma^4}.
  \end{align*}
  Thus, we derive the reference prior
  $\prior{\mu, \sigma^2} = \inv{\sigma^2}$.
  Following Example~\ref{ex:meanvar-cov}, we compute
  \begin{align*}
    \int_0^t \int_{-\infty}^\infty \post{\mu, \sigma^2} d \mu d\sigma^2 = 
    \inv{\Gamma(\divtwo{n-1})}\Gamma(\divtwo{n-1}, \inv{2 t}\pa{\y' \y - n \bar{y}^2}).
  \end{align*}
  I reran the coverage simulation from Example~\ref{ex:meanvar-cov} with this CDF and 
  got the results in Table~\ref{tab:normal-var-ref-cov}. Comparing to 
  Table~\ref{tab:normal-var-jeff-cov2}, we can see that the reference prior approach gives better
  results.
\end{example}
\begin{table}
\centering
\begin{tabular}{lccccccl}\toprule
\multicolumn{2}{c}{$n = 5$} & 
\multicolumn{2}{c}{$n = 10$} &
\multicolumn{2}{c}{$n = 15$} &
\multicolumn{2}{c}{$n = 20$}
\\\cmidrule(lr){1-2}
\cmidrule(lr){3-4}
\cmidrule(lr){5-6}
\cmidrule(lr){7-8}
$\sigma^2$ & $\textrm{coverage}$ & 
$\sigma^2$ & $\textrm{coverage}$ &
$\sigma^2$ & $\textrm{coverage}$ &
$\sigma^2$ & $\textrm{coverage}$ 
\\\midrule
0.1 & 0.9533 & 0.1 & 0.948 & 0.1 & 0.9504 & 0.1 & 0.9519 \\
0.5 & 0.9528 & 0.5 & 0.9499 & 0.5 & 0.9524 & 0.5 & 0.9486 \\
1.0 & 0.948 & 1.0 & 0.9503 & 1.0 & 0.9507 & 1.0 & 0.9484 \\
2.0 & 0.9529 & 2.0 & 0.9504 & 2.0 & 0.9515 & 2.0 & 0.9487 \\
5.0 & 0.9525 & 5.0 & 0.9507 & 5.0 & 0.9484 & 5.0 & 0.9511
\\\bottomrule
\end{tabular}
\caption{Frequentist coverages for the variance of a normal distribution with unknown mean and reference prior.}
\label{tab:normal-var-ref-cov}
\end{table}

\section{Noninformative Priors for Spatial Models}\label{sec:gpprior}
Let's consider noninformative priors for the Gaussian process~\eqref{eqn:gp}.
\begin{itemize}
  \item Using a constant prior isn't a viable option. In addition to the problem of incoherence,
    the resulting posterior would be improper \citep{objbay}. We might consider truncating the parameter space to
    make the constant prior proper, but that doesn't solve the problem as inference would be
    highly dependent on the truncation bounds.
  \item Certain modified forms of Jeffreys prior result in a proper posterior, but the credible 
    sets produced from the priors perform poorly \citep{rengp}.
\end{itemize}
That brings us to the reference prior approach. Since the model has multiple parameters,
we'll first integrate out $\rg$ and $\var$ using the conditional prior
$\prior{\rg, \var\gv\l,\nr} \propto \inv{\var}$.
Likelihood for Gaussian process~\eqref{eqn:gp} is given by
\begin{align*}
  \like{\rg, \var, \l, \nr} \propto
    \pa{\var}^{-n/2} \det{\G}^{-1/2}
    \exp\cpa{-\inv{2 \var} \multxt{\pa{\y - \X \rg}}{\G^{-1}}}
\end{align*}
where
  $\X = \pa{\x(\s_1), \ldots, \x(\s_n)}'$,
  $\G = \eta \id + \K(\l)$, and
  $\K(\l)_{ij} = \k\pa{\norm{\s_i - \s_j}}$.
Integrating likelihood with the conditional prior gives us
\begin{align}
  \likei{\l,\nr} &\propto \int_0^\infty \int_{\Re^p} \like{\rg, \var, \l, \nr} \prior{\rg, \var\gv\l, \nr} d\rg d\var \nonumber\\
                 &\propto \int_0^\infty \pa{\var}^{-(n-p)/2} \det{\G}^{-1/2} 
                 \det{\X' \Gi \X}^{-1/2}
                 \exp\cpa{-\frac{S^2}{2 \var}} \pa{\inv{\var}} d\var \nonumber\\
                 &\propto \det{\G}^{-1/2} \det{\X' \Gi \X}^{-1/2} \pa{S^2}^{-(n - p)/2} \label{eqn:likei}
\end{align}
where
\begin{align}
  S^2 = \multxt{\y}{\R}\and\quad
  \R = \Gi - \Gi \X \pa{\X'\Gi\X}^{-1} \X' \Gi. \label{eqn:R}
\end{align}
After computing the Fisher information matrix for $\likei{\cdot}$ and forming its Jeffrey prior, we
derive the complete prior
\begin{align}
  \prior{\rg, \var, \l, \nr} \propto \pa{\inv{\var}} \det{\smat(\l, \nr)}^{1/2} \label{eqn:prior}
\end{align}
where
\begin{equation}
  \smat(\l, \nr) = 
    \begin{pmatrix}
      \tr\cpa{(\R \dK)^2} & \tr(\R^2 \dK) & \tr(\R \dK) \\[6pt]
      * & \tr(\R^2) & \tr(\R) \\[6pt]
      * & * & n - p \\[6pt]
    \end{pmatrix}\label{eqn:smat}.
\end{equation}
For a detailed derivation, see \cite{rengp}.

To test the performance of the reference prior, we'll run the same simulations used in
\cite{rengp}. Details of how to compute the integrals will be given in \S\ref{sec:detalgo}.
\begin{example}\label{ex:gp-cov1}[\href{https://github.com/rnburn/bbai/blob/master/test/gaussian_process_prior_coverage.ipynb}{source}]
  To generate observations, I sample Gaussian process~\eqref{eqn:gp} with
  \begin{align*}
    \var = 1, \quad x_1(\s) = 1, \quad\rg_1 = 1, \and \quad\k(d) = \exp\cpa{-\frac{d}{\l}}
  \end{align*}
  at $10\times10$ evenly spaced points on the interval $[0, 1]\times[0, 1]$.
  I ran Algorithm~\ref{alg:covfreq} with $N = 200$ and allowed $\l$ and $\nr$ to vary. The
  results are given in Table~\ref{tab:gp-p1-cov}.
\end{example}
\begin{table}
\centering
\begin{tabular}{lcccccc}\toprule
& \multicolumn{3}{c}{$\nr = 0.01$} & 
\multicolumn{3}{c}{$\nr = 0.05$}
\\\cmidrule(lr){2-4}
\cmidrule(lr){5-7}
& $\l = 0.2$ & $\l = 0.5$ & $\l = 1.0$
& $\l = 0.2$ & $\l = 0.5$ & $\l = 1.0$
\\\midrule
  $\l$ coverage & 0.945 & 0.985 & 0.995 & 0.950 & 0.990 & 1.000 \\
  $\nr$ coverage & 0.885 & 0.980 & 0.995 & 1.000 & 0.995 & 1.000 \\
  $\var$ coverage & 0.990 & 0.995 & 0.980 & 0.975 & 0.985 & 0.985 \\
  $\beta_1$ coverage & 1.000 & 0.990 & 0.965 & 0.995 & 0.995 & 0.945
\\\midrule
& \multicolumn{3}{c}{$\nr = 0.1$} & 
\multicolumn{3}{c}{$\nr = 0.2$}
\\\cmidrule(lr){2-4}
\cmidrule(lr){5-7}
& $\l = 0.2$ & $\l = 0.5$ & $\l = 1.0$
& $\l = 0.2$ & $\l = 0.5$ & $\l = 1.0$
\\\midrule
  $\l$ coverage & 0.965 & 0.975 & 0.995 & 0.985 & 1.000 & 1.000 \\
  $\nr$ coverage & 1.000 & 0.975 & 0.995 & 0.995 & 0.970 & 0.990 \\
  $\var$ coverage & 1.000 & 0.985 & 0.985 & 0.970 & 0.985 & 0.980 \\
  $\beta_1$ coverage & 0.995 & 0.980 & 0.955 & 0.995 & 0.985 & 0.930
\\\bottomrule
\end{tabular}
\caption{Frequentist coverages for Gaussian process parameters on simulation data sets with a constant
regressor.}
\label{tab:gp-p1-cov}
\end{table}
\begin{example}
  [\href{https://github.com/rnburn/bbai/blob/master/test/gaussian_process_prior_coverage.ipynb}{source}]
  For the next simulation, I modify the Gaussian process in Example~\ref{ex:gp-cov1} to
  include additional regressors,
  $\x((u, v)) = \pa{1, u, v, u^2, u v, v^2}'$ with
    $\rg = \pa{0.15, -0.65, -0.1, 0.9, -1.0, 1.2}'$.
  Rerunning the simulation experiment with the same values of $\l$ and $\nr$ gave the
  coverages in Table~\ref{tab:gp-p6-cov}.
\end{example}
\begin{table}
\centering
\begin{tabular}{lcccccc}\toprule
& \multicolumn{3}{c}{$\nr = 0.01$} & 
\multicolumn{3}{c}{$\nr = 0.05$}
\\\cmidrule(lr){2-4}
\cmidrule(lr){5-7}
& $\l = 0.2$ & $\l = 0.5$ & $\l = 1.0$
& $\l = 0.2$ & $\l = 0.5$ & $\l = 1.0$
\\\midrule
  $\l$ coverage & 0.995 & 1.000 & 0.960 & 1.000 & 1.000 & 0.910 \\
  $\nr$ coverage & 0.865 & 0.950 & 0.915 & 1.000 & 1.000 & 0.990 \\
  $\var$ coverage & 0.995 & 0.975 & 0.835 & 1.000 & 0.985 & 0.760 \\
  $\beta_1$ coverage & 1.000 & 0.925 & 0.765 & 0.960 & 0.915 & 0.775 \\
  $\beta_2$ coverage & 0.945 & 0.895 & 0.870 & 0.935 & 0.900 & 0.845 \\
  $\beta_3$ coverage & 0.990 & 0.885 & 0.850 & 0.960 & 0.915 & 0.895 \\
  $\beta_4$ coverage & 0.915 & 0.900 & 0.835 & 0.940 & 0.890 & 0.855 \\
  $\beta_5$ coverage & 0.970 & 0.860 & 0.890 & 0.970 & 0.935 & 0.870 \\
  $\beta_6$ coverage & 0.935 & 0.900 & 0.840 & 0.955 & 0.910 & 0.875
\\\midrule
& \multicolumn{3}{c}{$\nr = 0.1$} & 
\multicolumn{3}{c}{$\nr = 0.2$}
\\\cmidrule(lr){2-4}
\cmidrule(lr){5-7}
& $\l = 0.2$ & $\l = 0.5$ & $\l = 1.0$
& $\l = 0.2$ & $\l = 0.5$ & $\l = 1.0$
\\\midrule
  $\l$ coverage & 1.000 & 1.000 & 0.890 & 1.000 & 1.000 & 0.815 \\
  $\nr$ coverage & 0.990 & 1.000 & 1.000 & 0.990 & 1.000 & 1.000 \\
  $\var$ coverage & 0.990 & 0.980 & 0.835 & 0.985 & 0.975 & 0.835 \\
  $\beta_1$ coverage & 0.980 & 0.870 & 0.795 & 0.960 & 0.900 & 0.800 \\
  $\beta_2$ coverage & 0.925 & 0.910 & 0.895 & 0.965 & 0.925 & 0.850 \\
  $\beta_3$ coverage & 0.970 & 0.915 & 0.865 & 0.940 & 0.900 & 0.905 \\
  $\beta_4$ coverage & 0.940 & 0.920 & 0.900 & 0.940 & 0.915 & 0.885 \\
  $\beta_5$ coverage & 0.945 & 0.870 & 0.895 & 0.960 & 0.865 & 0.870 \\
  $\beta_6$ coverage & 0.965 & 0.885 & 0.860 & 0.940 & 0.925 & 0.940
\\\bottomrule
\end{tabular}
\caption{Frequentist coverages for Gaussian process parameters on simulation data sets with polynomial
regressors.}
\label{tab:gp-p6-cov}
\end{table}
To test prediction performance, we can use a modified form of Algorithm~\ref{alg:covfreq}.
\begin{algorithm}[H]
  \caption{Test accuracy of prediction credible sets produced with a prior \label{alg:pcovfreq}}
  \begin{algorithmic}[1]
    \Statex
    \Function{prediction-coverage-test}{$\tilde{\params}$, $\alpha$}
      \Let{$cnt$}{$0$}
      \Let{$N$}{a large number}
      \For{$i \gets 1 \textrm{ to } N$}
      	\Let{$\tilde{\y}$}{sample from $P(\cdot\gv \tilde{\params})$}
        \Let{$t$}{
        $\int_{-\infty}^{\tilde{y}_1} \int\pr{y'\gv\params} \pi(\params\gv\tilde{y}_2, \ldots, \tilde{y}_n) d\params dy'$}
	\If{$\frac{\alpha}{2} < t < 1 - \frac{\alpha}{2}$}
          \Let{$cnt$}{$cnt + 1$}
        \EndIf
      \EndFor
      \State \Return{$\frac{cnt}{N}$}
    \EndFunction
  \end{algorithmic}
\end{algorithm}
\begin{example}\label{ex:gp-pcov}
  [\href{https://github.com/rnburn/bbai/blob/master/test/gaussian_process_prediction_coverage.ipynb}{source}]
  To generate observations, I sample from Gaussian process~\eqref{eqn:gp} with
  $\var = 1$ and $\k(d) = \exp\cpa{-\frac{d^2}{2\l^2}}$.
  I sampled training observations at $20$ evenly spaced points on the interval
  $[0, 1]$ and test observations at random points on the interval $[0, 1]$.
  I ran Algorithm~\ref{alg:pcovfreq} with $N = 100$ and varied $\l$ and $\nr$. 
  Table~\ref{tab:gp-pred-cov} shows
  the coverage results for Bayesian prediction distributions using the reference prior and
  maximum likelihood prediction distributions.
\end{example}
\begin{table}
\centering
\begin{tabular}{lcccccc}\toprule
& \multicolumn{3}{c}{$\nr = 0.001$} & 
\multicolumn{3}{c}{$\nr = 0.01$}
\\\cmidrule(lr){2-4}
\cmidrule(lr){5-7}
& $\l = 0.1$ & $\l = 0.2$ & $\l = 0.5$
& $\l = 0.1$ & $\l = 0.2$ & $\l = 0.5$
\\\midrule
  Bay coverage & 0.919 & 0.951 & 0.942 & 0.939 & 0.953 & 0.944 \\
  ML coverage & 0.812 & 0.905 & 0.934 & 0.838 & 0.912 & 0.919
\\\midrule
& \multicolumn{3}{c}{$\nr = 0.1$} & 
\multicolumn{3}{c}{$\nr = 0.2$}
\\\cmidrule(lr){2-4}
\cmidrule(lr){5-7}
& $\l = 0.1$ & $\l = 0.2$ & $\l = 0.5$
& $\l = 0.1$ & $\l = 0.2$ & $\l = 0.5$
\\\midrule
  Bay coverage & 0.929 & 0.943 & 0.932 & 0.936 & 0.937 & 0.938 \\
  ML coverage & 0.847 & 0.893 & 0.920 & 0.853 & 0.893 & 0.903
\\\bottomrule
\end{tabular}
\caption{Frequentist coverages for Bayesian and maximum likelihood Gaussian process predictions on simulation data sets.}
\label{tab:gp-pred-cov}
\end{table}

\section{Deterministic Bayesian Inference}\label{sec:detalgo}
The key component for deterministic prediction and inference is an 
accurate approximation to the posterior distribution for $\l$  and $\nr$ that enables
efficient computation of integrals,
$\postx{\l, \nr} \approx \likei{\l, \nr} \times \prior{\l, \nr}$
where
$\prior{\l, \nr} \propto \det{\smat(\l, \nr)}^{1/2}$
and $\smat(\cdot)$ is defined in \eqref{eqn:smat}.

Given $\postx{\cdot}$, it's relatively straightforward to derive
approximations for the marginal distributions
\begin{align*}
  \post{\l} &\approx \int_0^\infty \postx{\l, \nr} d\nr, \\
  \post{\nr} &\approx \int_0^\infty \postx{\l, \nr} d\l, \and \\
  \post{\var} &\approx \int_0^\infty \int_0^\infty \prp{\var\gv\y,\l,\nr} \postx{\l, \nr} d\l d\nr
\end{align*}
and approximations for prediction distributions,
\begin{align*}
  \prp{Z(\s)\gv\y} &\approx \int_0^\infty \int_0^\infty \prp{Z(\s)\gv\y,\l,\nr}\postx{\l, \nr} d\l d\nr.
\end{align*}
\subsection*{Outline of Algorithm}
Assume $\varphi_\l(\cdot)$ and $\varphi_\nr(\cdot)$ are strictly increasing
functions onto $(0, \infty)$ with continuous derivatives. Put
\begin{align}
  f(\u) = \begin{multlined}[t]
    -\log\likei{\varphi_\l(u_1), \varphi_\nr(u_2)} \\- \log\prior{\varphi_\l(u_1), \varphi_\nr(u_2)} - 
    \log \dot{\varphi}_\l(u_1) - \log \dot{\varphi_\nr}(u_2).
  \end{multlined}\label{eqn:obj}
\end{align}
$f(\cdot)$ is the negative log of the reparameterized posterior $\post{\l, \nr}$.
Approximation of $\exp(-f(\cdot))$ naturally leads to approximation and integration of 
$\post{\l, \nr}$.

We'll build an approximation in four steps.
\begin{algorithm}[H]
  \caption{Build a multivariate polynomial to approximate $\exp(-f(\cdot))$ where $f(\cdot)$ \eqref{eqn:obj} 
    is the negative negative log of the reparameterized posterior
  function $\post{\l, \nr}$\label{alg:high}}
  \begin{algorithmic}[1]
    \State\label{step:opt} Using a trust-region optimizer and exact equations for $\nabla f$ and $\nabla^2 f$,
    minimize $f$ to find $\umap$.

  \State\label{step:bracket} Let $\v_1$ and $\v_2$ denote two orthonormal eigenvectors of the Hessian at $\umap$,
    $\nabla^2 f(\umap)$.
    Find values $a_1 < 0 < b_1$ and $a_2 < 0 < b_2$ such that
    \begin{align*}
      -\pa{f(\umap + a_i \v_i) - f(\umap)} &= \log \varepsilon_1(a_i) \and \\
      -\pa{f(\umap + b_i \v_i) - f(\umap)} &= \log \varepsilon_2(b_i)
    \end{align*}
    for $i=1,\ 2$ and $\varepsilon_i(\cdot)$ small.
    These values bracket $f(\cdot)$ around a rectangular region oriented
    along the eigenvectors $\v_1$ and $\v_2$ that contains most of the probability mass.

  \State\label{step:warp} Find monotonic cubic splines $\wrp_1(\cdot)$ and $\wrp_2(\cdot)$ such that
      $\wrp_i(0) = a_i$, $\wrp_i(0.5) = 0$, and $\wrp_i(1) = b_i$
    for $i=1,2$.

  \State\label{step:sg} Put
    \begin{align}
      g(\x) = \exp\cpa{-\pa{f\pa{\umap + \wrp_1(x_1) \v_1 + \wrp_2(x_2)\v_2} - f(\umap)}}. \label{eqn:interf}
    \end{align}
    Using Chebyshev nodes and the eigenvectors $\v_1$ and $\v_2$ for a basis, adaptively
    build a sparse grid and interpolating polynomial to approximate $g(\cdot)$ (and hence 
    $\post{\l, \nr}$) over the region $[0, 1]\times[0,1]$.
  \end{algorithmic}
\end{algorithm}
Proposition~7 and Proposition~9 from \cite{rengp} show that $\post{\l, \nr}$ is bounded as
$\l\to0$ or $\nr\to0$ and derive $\bigo{\cdot}$ functions for when $\l\to\infty$ and $\nr\to\infty$.
Using suitable choices of $\varphi_\l(\cdot)$, $\varphi_\nr(\cdot)$, and $\varepsilon_i(\cdot)$, we 
can achieve bounds for the probability mass outside of the bracketing region in Step~\ref{step:bracket}.
Following \cite{postparam}, we use the parameterization
$\phi_\l(t) = \phi_\nr(t) = \exp(t)$.
We'll only consider the simple case of $\varepsilon_i$ fixed to some small constant, but other choices could lead to tighter bounding. 

We can use any decent root-finding algorithm (e.g., Newton's method) for Step~\ref{step:bracket};
we use the monotonic cubic algorithm from \cite{monotonic-cubic} for Step~\ref{step:warp}. 
Step~\ref{step:opt} and Step~\ref{step:sg} are more complicated, and I break them down in greater detail in the next sections.
\subsection*{Step~\ref{step:opt}: Trust-region Optimization}
Let $f\colon \mathbb{R}^p \to \mathbb{R}$ denote a twice-differentiable objective function. 
Trust-region methods are iterative, second-order optimization algorithms that produce a sequence
$\left\{\bs{x}_k\right\}$ where the $k$\ts{th} iteration is generated by updating the previous iteration
with a solution to the subproblem \citep{sorensen:82}
\begin{align*}
  \x_k &= \x_{k-1} + \s_k \and \\
  \s_k &= \argmin_\s \cpa{
      \nabla f(\x_{k-1})' \s + 
      \half \s' \nabla^2 f(\x_{k-1}) \s
    } \\
   &\phantom{=} \quad\quad\textrm{such that} \; \norm{\s} \le \delta_k.
\end{align*}
The subproblem minimizes the second-order approximation of $f$ at $\bs{x}_{k-1}$ within the 
neighborhood $\| \bs{s} \| \le \delta_k$, called the trust region. Using the trust region, 
we can restrict the second-order approximation to areas where it models $f$ well. Efficient algorithms
exist to solve the subproblem regardless of whether $\nabla^2 f(\bs{x}_{k-1})$ is positive-definite,
making trust-region methods well-suited for non-convex optimization problems \citep{more:83}. 
With proper rules for updating $\delta_k$ and standard assumptions, such as Lipschitz continuity of 
$\nabla f$, trust-region methods are globally convergent. Moreover, if $\nabla^2 f$ is 
Lipschitz continuous for all $\bs{x}$ sufficiently close to a nondegenerate second-order stationary 
point $\bs{x}_*$ where $\nabla^2 f(\bs{x}_*)$  is positive-definite, then trust-region methods 
have quadratic local convergence \citep{nocedal:06}.

Algorithm~\ref{alg:opt} describes the trust-region algorithm we use for Step~\ref{step:opt}, and Appendix~\ref{appendix:a} derives 
equations for evaluating the value, gradient, and Hessian of the objective~\eqref{eqn:obj}.
\begin{algorithm}[H]
  \caption{Minimize an objective function $f(\cdot)$\label{alg:opt}}
  \begin{algorithmic}[1]
    \Function{minimize}{$f, \x_0$}
      \Let{$tol$}{tolerance}
      \Let{$\delta_0$}{an initial trust-region radius}
      \Let{$y_0$}{$f(\x_0)$}
      \Let{$\grad_0$}{$\nabla f(\x_0)$}
      \Let{$\hess_0$}{$\nabla^2 f(\x_0)$}
      \Let{$k$}{$0$}
      \While{$\norm{\grad_k}_\infty > tol$ or $\hess_k$ is not positive definite}
        \Let{$\x_{k+1}$, $y_{k+1}$, $\delta_{k+1}$}{\Call{compute-next-step}{$\x_k$, $y_k$, $\grad_k$, $\hess_k$, $\delta_k$}}
        \Let{$\grad_{k+1}$}{$\nabla f(\x_{k+1})$}
        \Let{$\hess_{k+1}$}{$\nabla^2 f(\x_{k+1})$}
        \Let{$k$}{$k + 1$}
      \EndWhile
      \State \Return{$\x_k$, $y_k$, $\hess_k$}
    \EndFunction
    \Function{compute-next-step}{$\x_k$, $y_k$, $\grad_k$, $\hess_k$, $\delta_k$} 
      \Let{$\delta_{k+1}$}{$\delta_k$}
      \While{$1$}
      \Let{$\s_k$}{$
        \argmin_\s \cpa{
          \grad_k' \s + \half \s' \hess_k \s\;
          \bigg|
          \norm{\s} \le \delta_{k+1}
        }
        $
      }

      \Comment{\parbox[t]{0.75\linewidth}{
          \strut Solve the trust-region subproblem \citep{more:83}}
        \strut
      }
      \Let{$\x_{k+1}$}{$\x_k + \s_k$}
      \Let{$y_{k+1}$}{$f(\x_{k+1})$}
      \Let{$\rho$}{$\frac{y_{k+1} - y_k}{\grad_k' \s_k + \half\s_k'\hess_k\s_k}$}

      \Comment{\parbox[t]{0.75\linewidth}{
          \strut $\rho$ measures the accuracy of the second-order Taylor approximation to $f(\cdot)$ 
          within the trust-region neighborhood, $\delta_{k+1}$, about $\x_k$
        \strut
      }}
      \If{$\rho < \frac{1}{4}$}
      \Let{$\delta_{k+1}$}{$\inv{4}\delta_{k+1}$}\linecomment{Shrink the trust region}
      \ElsIf{$\rho > \frac{3}{4}$ and $\norm{\s_k} = \delta_{k+1}$}
        \Let{$\delta_{k+1}$}{$2\delta_{k+1}$}\linecomment{Expand the trust region}
      \EndIf
      \If{$\rho > \frac{1}{4}$}
        \State \Return{$\x_{k+1}$, $y_{k+1}$, $\delta_{k+1}$}
      \EndIf
      \EndWhile
    \EndFunction
  \end{algorithmic}
\end{algorithm}

\subsection*{Step~\ref{step:sg}: Sparse Grid Approximation}
We seek to approximate $g(\cdot)$ \eqref{eqn:interf} by a polynomial $\tilde{g}(\cdot)$ that interpolates
$g(\cdot)$ at points in $[0, 1]\times[0, 1]$. If we choose the points well, we can achieve high accuracy
with a minimal number of points, making $\tilde{g}(\cdot)$ cheaper to build and evaluate.

The simplest approach would be to interpolate at equispaced points, but polynomials at equispaced 
points perform terribly (see Runge's phenomenon). Much better is to interpolate at Chebyshev nodes.
Polynomials at Chebyshev nodes have excellent approximation performance \citep{approxbook}, but interpolating
on a dense grid would still be expensive. We can achieve better efficiency if we interpolate on a sparse grid,
and we can achieve even better efficiency if we adaptively construct the sparse grid to avoid 
unnecessary evaluations in areas that can be approximated well by lower-order polynomials. 

Put
\begin{align*}
  X^i &= \cpa{x_1^i, \ldots, x_{m_i}^i}, \\
  m_i &= 
  \begin{cases}
    1 & \text{if}\ i = 0, \\
    2^{i-1} + 1 & \text{otherwise},
  \end{cases} \\
  x_j^i &= 
    \begin{cases}
      \half &\text{if}\ i=0, \\
      \frac{1}{2}\pa{1 - \cos \frac{\pi(j-1)}{m_i-1}} &\text{otherwise}.
    \end{cases}
\end{align*}
The Chebyshev-Gauss-Lobatto nodes, $\cpa{X^i}$, form a nested sequence
of points, $X^i \subset X^{i+1}$,
that serve as a building block for constructing interpolations and quadrature
rules for sparse grids \citep{cheby1, cheby2}.
Let $\psi_j^i(\cdot)$ denote the unique $(m_i-1)$-degree polynomial where
\begin{align*}
  \psi_j^i(x_{j'}^i) = 
    \begin{cases}
      1 & \text{if}\ j = j', \\
      0 & \text{otherwise};
    \end{cases}
\end{align*}
let $V^i$ denote the vector space spanned by the basis functions
$\cpa{\psi_{j}^{i}}$ for $j=1, \ldots, m_i$;
and define 
  $\Delta V^0 = V^0$,
  $\Delta V^i = V^i - V^{i-1}$ for $i>0$.
We will build an approximation using functions from vector spaces
\begin{align*}
  W^\iset = \bigoplus_{\iv\in \iset} \Delta V^{i_1} \otimes \cdots \otimes \Delta V^{i_d}
\end{align*}
where the index set $\iset$ is required to be \textit{admissible}:
if $\iv \in \iset$ and $i_k>0$, then $\iv - \ev_k \in \iset$. The vector spaces
$W^\iset$ are
a generalization of Smolyak sparse grids and allow for different dimensions to
have different levels of refinement \citep{sgdim}.

To build the sparse grid, we follow the algorithm from \cite{chebyadapt} and greedily 
add indexes and nodes with the largest approximation errors until a target accuracy is
achieved. The algorithm adapts by
both dimension and locality.

Define
  $\x_\jv^\iv = \pa{x_{j_1}^{i_1}, \ldots, x_{j_d}^{i_d}}$,
  $\psi_\jv^\iv(\x) = \psi_{j_1}^{i_1}(x_1) \cdots \psi_{j_d}^{i_d}(x_d)$,
  $\Delta X^0 = X^0$, and $\Delta X^i = X^i \setminus X^{i-1}$.
\begin{algorithm}[H]
  \caption{Build an interpolating polynomial on a sparse grid to approximate a function $f(\cdot)$ on $[0, 1]^d$ \label{alg:sg}. Continue to refine
  the sparse grid until errors, specified by a function $\ferr(\cdot)$, are within a target tolerance.}
  \begin{algorithmic}[1]
    \Function{approximate}{$f$, $\ferr$}
      \Let{$tol$}{tolerance}
      \Let{$G$}{$\cpa{}$}\linecomment{Subgrids and surpluses for the sparse grid}
      \Let{$\fringe$}{$\cpa{}$}\linecomment{Expanded subgrids not yet added to $G$}
      \Let{$\iv$}{$\nullv$}
      \Let{$\fringe$}{$\fringe\cup\{$\Call{expand-subgrid}{$G$, $f$, $\ferr$, $\iv$}$\}$} 
      \While{$1$}
        \Let{$\iv$, $\err_\jv^\iv$}{pick $\iv$, $\err_\jv^\iv$ to maximize $\err_\jv^\iv$ in $\fringe$}
        \If{$\err_\jv^\iv < tol$}
          \State \Return{$G$}
        \EndIf
        \Let{$G$}{$G\cup\cpa{(\iv, \z^\iv, \errs^\iv)}$}
        \Let{$\fringe$}{$\fringe\setminus\cpa{(\iv, \z^\iv, \errs^\iv)}$}
        \For{$\ifwd$ in $\cpa{\iv + \ev_k \mid 1 \le k \le d}$}
        \If{for all $k$ such that $(\ifwd)_k > 0$, $\ifwd - \ev_k$ is in $G$}
          \Let{$\fringe$}{$\fringe\cup\{$\Call{expand-subgrid}{$G$, $f$, $\ferr$, $\ifwd$}$\}$} 
        \EndIf
        \EndFor
      \EndWhile
    \EndFunction
    \Function{expand-subgrid}{$G$, $f$, $\ferr$, $\iv$}

      \Comment{\parbox[t]{0.85\linewidth}{
        \strut
        Compute surpluses and errors for every active refinement node of the
        subgrid defined by $\iv$
        \strut
      }
      }
      \Let{$\z^\iv$}{$\nullv$}
      \Let{$\errs^\iv$}{$\nullv$}
      \For{$\x_\jv^\iv$ in $\Delta X^{i_1} \otimes \cdots \otimes \Delta X^{i_d}$}
        \If{\Call{is-active}{$G$, $\iv$, $\jv$} or $\iv = \nullv$}
          \Let{$y$}{$f(\x_\jv^\iv)$}
          \Let{$\tilde{y}$}{\Call{evaluate}{$G$, $\x_\jv^\iv$}}
          \Let{$z_\jv^\iv$}{$y - \tilde{y}$}
          \Let{$\err_\jv^\iv$}{$\ferr(y, \tilde{y}, \x_\jv^\iv)$}
        \EndIf
      \EndFor
      \State \Return{$\pa{\iv, \z^\iv, \errs^\iv}$}
    \EndFunction
    \algstore{approximate}
  \end{algorithmic}
\end{algorithm}
\addtocounter{algorithm}{-1}
\begin{algorithm}[H]
  \caption{(continued)}
  \begin{algorithmic}[1]
    \algrestore{approximate}
    \Function{evaluate}{$G$, $\x$}

      \Comment{\parbox[t]{0.85\linewidth}{
        \strut
        Evaluate at $\x$ the polynomial interpolating the sparse grid $G$
        \strut
      }
      }
      \Let{$res$}{$0$}
      \For{$\iv, \z^\iv$ in $G$}
      \Let{$res$}{$
        res + \sum_{z_\jv^\iv \in \z^\iv} 
        z_\jv^\iv \psi_\jv^\iv(\x)
      $}
      \EndFor
      \State \Return{$res$}
    \EndFunction
    \Function{is-active}{$G$, $\iv$, $\jv$}
      
      \Comment{\parbox[t]{0.85\linewidth}{
          \strut Determine if the refinement $x_{\jv}^{\iv}$ is \textit{active}. A refinement
        is active if it has at least one neighbor with an error that exceeds the cutoff threshold.
        \strut
      }
      }
      \Let{$\tau$}{cutoff threshold}
      \For{$k$ such that $i_k > 0$}
        \Let{$\ibwd$}{$\iv - \ev_k$}
        \For{$\err_\jbwd^\ibwd$ in $G$}
          \If{$\err_\jbwd^\ibwd > \tau$ and \Call{is-point-neighbor}{$\ibwd$, $\jbwd$, $\jv$, $k$}}
            \State\Return{$1$}
          \EndIf
        \EndFor
      \EndFor
      \State\Return{$0$}
    \EndFunction
    \Function{is-point-neighbor}{$\iv$, $\jv$, $\jv'$, $k$}

      \Comment{\parbox[t]{0.85\linewidth}{
        \strut
        Determine if the refinement $x_{\jv'}^{\iv + \ev_k}$ neighbors the point
      $x_{\jv}^{\iv}$. See \S4.2 of \cite{chebyadapt} for details.
        \strut
    }
      }
      \If{there exists $k' \ne k$ such that $j_{k'} \ne j'_{k'}$}
        \State\Return{$0$}
      \ElsIf{$i_k \le 1$}
        \State\Return{$1$}
      \Else
        \State\Return{
          $\pa{x_{j_k-1}^{i_k} < x_{j'_k}^{i_k+1} < x_{j_k}^{i_k}}$ or
          $\pa{x_{j_k}^{i_k} < x_{j'_k}^{i_k+1} < x_{j_k+1}^{i_k}}$
        }
      \EndIf
    \EndFunction
  \end{algorithmic}
\end{algorithm}
At $\umap$, a second-order Taylor approximation to $f(\cdot)$ \eqref{eqn:obj} gives us
\begin{align*}
  f(\umap + \bs{\delta}) \approx
  f(\umap) + \half \bs{\delta}'(\nabla^2 f)(\umap)\bs{\delta}.
\end{align*}
If we use the eigenvectors $\v_1$ and $\v_2$ as a basis, we have
\begin{align*}
  f(\umap + \delta_1 \v_1 + \delta_2 \v_2) \approx
  f(\umap) + \half \pa{\xi_1 \delta_1^2 + \xi_2 \delta_2^2}
\end{align*}
where $\xi_1$ and $\xi_2$ are the eigenvalues of $\nabla^2f(\umap)$. Thus,
$\exp(f(\cdot))$ is approximately separable at $\umap$ along the eigenvectors; and hence, $g(\cdot)$ \eqref{eqn:interf} is
approximately separable at $\bs{0.5}$,
\begin{align*}
  g(0.5 + \delta_1, 0.5 + \delta_2) \approx h_1(0.5 + \delta_1) \times h_2(0.5 + \delta_2)
\end{align*}
for some $h_d$ and $\delta_d$ small.
We can use this observation to build a more efficient approximation. Let $\tilde{h}_1(\cdot)$ and
$\tilde{h}_2(\cdot)$ denote interpolations at Chebyshev nodes of the functions 
$g(\cdot, 0.5)$ and $g(0.5, \cdot)$. Then run Algorithm~\ref{alg:sg} with
the target function $\frac{g(x_1, x_2)}{\tilde{h}_1(x_1)\tilde{h}_2(x_2)}$ and
the error function $\ferr(y, \tilde{y}, x_1, x_2) = \left\lvert (y - \tilde{y}) \tilde{h}_1(x_1)\tilde{h}_2(x_2) \right\rvert$.

\begin{figure}
  \centering
  \includegraphics{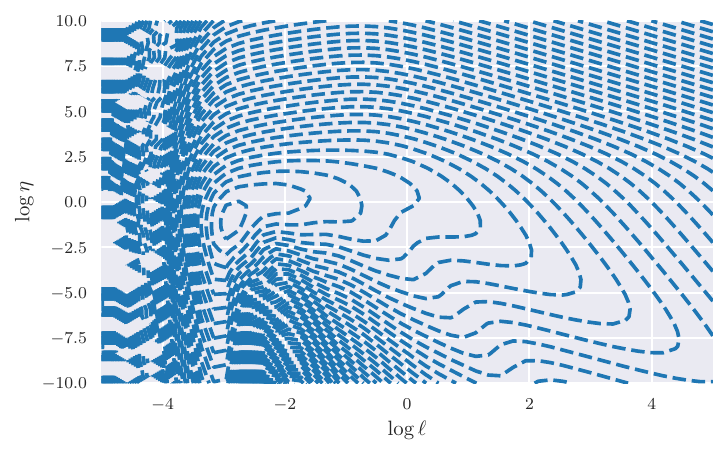}
  \caption{Reparameterized log posterior for the Example~\ref{ex:predictgpml} data set with reference prior}
\label{fig:ex1-post}
\end{figure}

\begin{figure}
  \centering
  \includegraphics{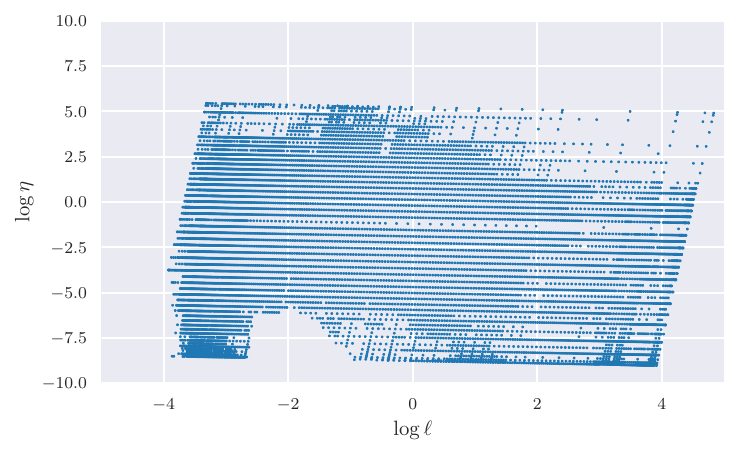}
  \caption{Sparse grid used to interpolate the reparameterized posterior for the Example~\ref{ex:predictgpml} data set with reference prior}
\label{fig:ex1-sparse-grid}
\end{figure}

\begin{example} (Example~\ref{ex:predictgpbay} continued)
I ran Algorithm~\ref{alg:sg} on the data set from Example~\ref{ex:predictgpml}. Figure~\ref{fig:ex1-post}
shows contours for the log of the reparameterized posterior function, and 
Figure~\ref{fig:ex1-sparse-grid} shows the sparse grid used to approximate the reparameterized
posterior.
\end{example}

\subsection*{Prediction Distributions}
The sparse grid from Algorithm~\ref{alg:sg} naturally leads to
a quadrature rule to approximate integration \citep{chebyadapt}. 
Let $f(\l, \nr)$ denote a function. Put
$\bs{\varphi}(\u) = \pa{\varphi_\l(u_1), \varphi_\nr(u_2)}'$. Then
\begin{align}
  \int_0^\infty \int_0^\infty &f(\l, \nr) \post{\l, \nr} d\l d\nr  \nonumber\\
      &\approx \int_0^\infty \int_0^\infty f(\l, \nr) \postx{\l, \nr} d\l d\nr \nonumber\\
      &\approx \!\begin{multlined}[t]\inv{Z} \int_0^1 \int_0^1 
        f\pa{\bs{\varphi}(\umap + \wrp_1(x_1) \v_1 + \wrp_2(x_2)\v_2)} \\
      g(x_1, x_2) \dwrp_1(x_1) \dwrp_2(x_2) dx_1 dx_2 \end{multlined}\nonumber\\
      &\approx \sum_k w_k f(\l_k, \nr_k) \label{eqn:postw}
\end{align}
where the points $\cpa{(\l_k, \nr_k)'}$ are the transformed nodes
of the sparse grid and weights are derived from integrals of the basis functions with the
separable approximations,
\begin{align*}
  \int_0^1 \psi_j^i(x) \tilde{h}_k(x) dx
\end{align*}
for $k=1,2$.

Let $\tilde{\s}$ denote unobserved locations. Then
\begin{align*}
  \prp{Z(\tilde{s}_1), \ldots, Z(\tilde{s}_m) \gv \y}
    \approx \sum_k w_k \prp{Z(\tilde{s}_1), \ldots, Z(\tilde{s}_m) \gv \y, \l_k, \nr_k}
\end{align*}
gives us an approximation of the prediction distribution. Let's derive a more
explicit formula for the conditional probability $\prp{\cdot \gv \y, \l, \nr}$. Use
$\y_1 = \y$ to denote the observations and use $\y_2$ to denote possible values at
the unobserved locations $\tilde{s}_1, \ldots, \tilde{s}_m$. 
Applying \eqref{eqn:likei}, we have
\begin{align*}
  \prp{\y_2\gv\y_1,\l,\nr} &\propto 
  \int_0^\infty \int_{\Re^p} \pr{\y_1, \y_2 \gv \rg, \var, \l, \nr} \pa{\inv{\var}} d\rg d\var \\
                           &\propto \bra{\pa{\y_1, \y_2} \R \pa{\y_1, \y_2}'}^{-(n + m - p)/2},
\end{align*}
where $\R$ is given by \eqref{eqn:R}. 
Put
$\R =
    \begin{pmatrix}
      \R_{11} & \R_{12} \\
      \R_{12}' & \R_{22}
    \end{pmatrix}$.
Then
\begin{align*}
  \pa{\y_1, \y_2} \R \pa{\y_1, \y_2}'
    &= \y_1' \R_{11} \y_1 + 2 \y_1' \R_{12} \y_2 + \y_2' \R_{22} \y_2 \\
    &= \pa{\y_2 - \bar{\y}_2}' \R_{22} \pa{\y_2 - \bar{\y}_2} + b
\end{align*}
where
$\bar{\y}_2 = -\R_{22}^{-1} \R_{12}' \y_1$ and
  $b = \y_1' \R_{11} \y_1 -\bar{\y}_2' \R_{22} \bar{\y}_2$.

\subsection*{$\var$ Marginal}
The marginal distribution of $\var$ is given by
\begin{align*}
\prp{\var\gv\y} 
  &= \int_0^\infty \int_0^\infty
        \prp{\var\gv\y, \l, \nr} \post{\l, \nr} d\l d\nr \\
  &\approx \sum_k w_k \prp{\var\gv\y, \l_k, \nr_k},
\end{align*}
where $\cpa{w_k}$, $\cpa{\l_k}$, and $\cpa{\nr_k}$ are defined in \eqref{eqn:postw}. From \eqref{eqn:likei}, we have
\begin{align}
  \prp{\var\gv\y, \l, \nr} 
  &\propto
  \int_{\Re^p} \like{\rg, \var, \l, \nr} \prior{\rg, \var\gv\l, \nr} d\rg \nonumber\\
  &\propto \pa{\var}^{-(n-p)/2} \det{\G}^{-1/2} 
                 \det{\X' \Gi \X}^{-1/2}
                 \exp\cpa{-\frac{S^2}{2 \var}} \pa{\inv{\var}} \nonumber\\
  &\propto \pa{\inv{\var}}^{(n - p)/2 + 1}
    \exp\cpa{-\frac{S^2}{2 \var}}. \label{eqn:vardist}
\end{align}
\eqref{eqn:vardist} is the unnormalized PDF of an inverse-gamma distribution. Normalizing
gives us
\begin{align*}
  \prp{\var\gv\y, \l, \nr} 
  = \frac{\pa{S^2/2}^{(n-p)/2}}{\Gamma((n-p)/2)}
  \pa{\inv{\var}}^{(n - p)/2 + 1}
    \exp\cpa{-\frac{S^2}{2 \var}}.
\end{align*}

\subsection*{$\rg$ Marginals}
Similarly, to compute the posterior distribution of a particular regressor $\beta_j$, we have
\begin{align*}
  \prp{\beta_j\gv\y} 
  &= \int_0^\infty \int_0^\infty
        \prp{\beta_j\gv\y, \l, \nr} \post{\l, \nr} d\l d\nr \\
\end{align*}
where
\begin{align*}
  \prp{\beta_j\gv\y, \l, \nr} 
  &\propto \begin{multlined}[t]
    \int_0^\infty \int_{\Re^{p-1}} \pa{\inv{\var}}^{n/2 + 1} \\
    \exp\cpa{-\inv{2 \var} \multxt{\pa{\y - \X \rg}}{\Gi}} d\rg_{/j} d\var
  \end{multlined} \\
  &\propto \begin{multlined}[t]
    \int_0^\infty \int_{\Re^{p-1}} \pa{\inv{\var}}^{n/2 + 1} \\
    \exp\cpa{-\inv{2 \var} \multxt{\pa{\rg - \bar{\rg}}}{\X'\Gi\X}} \\
    \exp\cpa{-\inv{2 \var} \pa{\y' \Gi \y - \bar{\rg} \X' \Gi \X \bar{\rg}}}
    d\rg_{/j} d\var
  \end{multlined} \\
  &\propto \begin{multlined}[t]
    \int_0^\infty \pa{\inv{\var}}^{(n - p + 1)/2 + 1}
    \exp\cpa{-\inv{2 \var} \inv{\pa{\Ai}_{jj}} \pa{\beta_j - \bar{\beta}_j}^2 } \\
    \exp\cpa{-\inv{2 \var} \pa{\y' \Gi \y - \bar{\rg} \X' \Gi \X \bar{\rg}}}
    d\var
  \end{multlined} \\
  &\propto \int_0^\infty \pa{\inv{\var}}^{(n - p + 1)/2 + 1}
    \exp\cpa{-\inv{2 \var} \#1} d\var \\
  &\propto \pa{\#1}^{-(n - p + 1)/2}
\end{align*}
and
\begin{align*}
  \A &= \X' \Gi \X \\
  \bar{\beta} &= \Ai \X' \Gi \y \\
  \#1 &= \inv{\pa{\Ai}_{jj}} \pa{\beta_j - \bar{\beta}_j}^2 + 
         \y' \Gi \y - 
         \bar{\rg}' \A \bar{\rg} \\
  &= \inv{\pa{\Ai}_{jj}} \pa{\beta_j - \bar{\beta}_j}^2 + S^2.
\end{align*}
We recognize $\prp{\beta_j\gv\y}$ as being a t-distribution with $n - p$ degrees of freedom,
mean $\rg_j$, and scale
$s_{\beta_j} = \cpa{\frac{\pa{\Ai}_{jj} S^2}{n - p}}^{1/2}$.

\subsection*{$\l$, $\nr$ Marginals}
Algorithm~\ref{alg:sg} gives us an interpolating function that is inexpensive to evaluate and accurately 
approximates the
reparameterized posterior
$\post{u_1, u_2} = \post{\varphi_\l(u_1), \varphi_\nr(u_2)} 
    \dot{\varphi}_\l(u_1) \dot{\varphi}_\nr(u_2)$.
Now,
\begin{align*}
  \post{u_1} &= \int \post{u_1, u_2} d u_2 \\
             &\approx \sum_k w_k \postx{u_1, t_k},
\end{align*}
where $\cpa{w_k}$ and $\cpa{t_k}$ can be chosen by the Gauss-Legendre quadrature rule for the interval
determined by the bracketing region in Step~\ref{step:bracket} of Algorithm~\ref{alg:high}. If we evaluate $\postx{u_1}$ at Chebyshev nodes
across the range of $u_1$ in the bracketing region, then we obtain a polynomial that approximates $\post{u_1}$.
$\post{u_2}$ can be similarly approximated using Chebyshev nodes.

If the error bounds from Step~\ref{step:bracket} are tight enough, the
polynomial approximations for $\post{u_1}$ and $\post{u_2}$ will be suitable
for estimating the CDFs. But since they cut off integration outside of the
bracketing region, they won't accurately capture endpoint behavior and shouldn't
be used for estimating moments. For example, $\post{\nr}$ has infinite mean,
which won't be reflected by the polynomial approximation.

\section{Real Data Analysis}
Let's apply the algorithms from \S\ref{sec:detalgo} to real data. 

\subsection{Soil Carbon-to-Nitrogen}\label{sec:soil}
\begin{figure}
  \begin{subfigure}[t]{\textwidth}
    \centering
    \scalebox{1.0}{\includegraphics{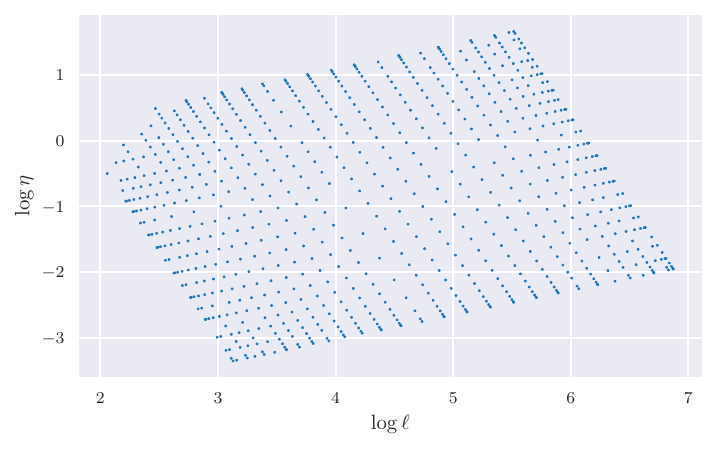}}
  \end{subfigure}
  \caption{Sparse grid used to interpolate the posterior of the soil data set}
  \label{fig:soil-sparse-grid}
\end{figure}

\begin{figure}
  \centering
  \begin{subfigure}[t]{0.45\textwidth}
    \centering
    \scalebox{0.45}{\includegraphics{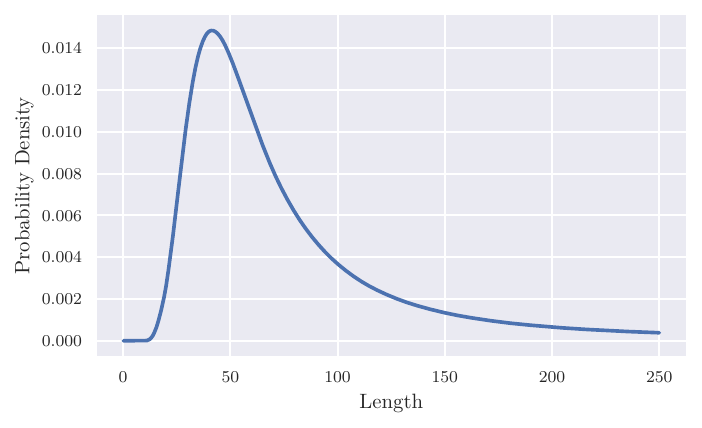}}
  \end{subfigure}%
  ~
  \begin{subfigure}[t]{0.45\textwidth}
    \centering
    \scalebox{0.45}{\includegraphics{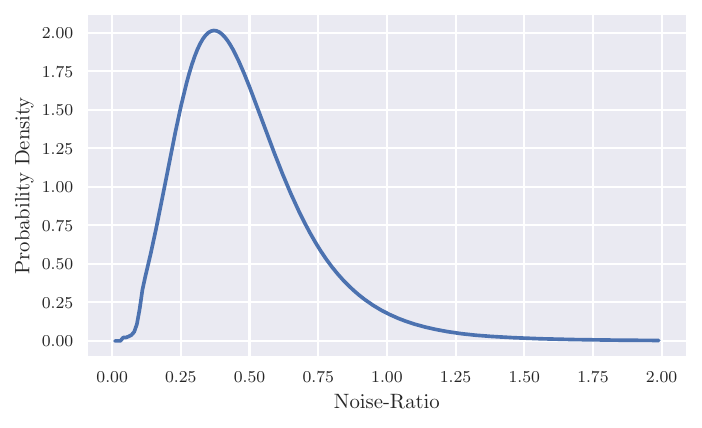}}
  \end{subfigure}
  \begin{subfigure}[t]{0.45\textwidth}
    \centering
    \scalebox{0.45}{\includegraphics{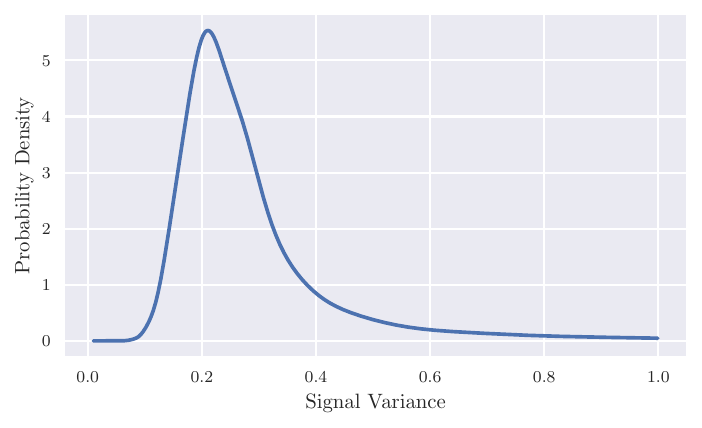}}
  \end{subfigure}%
  ~
  \begin{subfigure}[t]{0.45\textwidth}
    \centering
    \scalebox{0.45}{\includegraphics{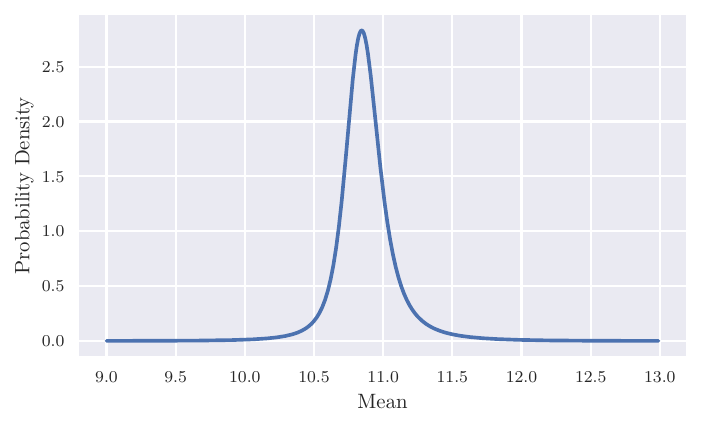}}
  \end{subfigure}
  \caption{Marginalizations of the posterior distribution of the soil data set}
  \label{fig:soil-marg}
\end{figure}

\begin{figure}
  \centering
  \begin{subfigure}[t]{0.45\textwidth}
    \centering
    \scalebox{0.45}{\includegraphics{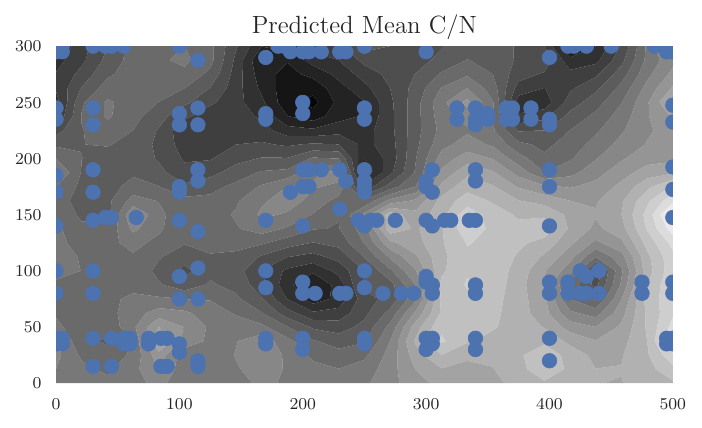}}
  \end{subfigure}%
  ~
  \begin{subfigure}[t]{0.45\textwidth}
    \centering
    \scalebox{0.45}{\includegraphics{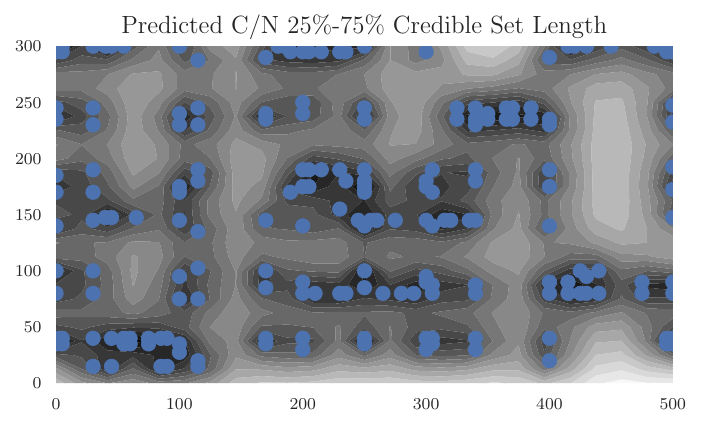}}
  \end{subfigure}
  \caption{Prediction means and credible sets of soil carbon-to-nitrogen ratios for soil data set with sampling 
  locations}
  \label{fig:soil-pred}
\end{figure}

We'll first look
at a data set from \cite{soil} of carbon-to-nitrogen ratios sampled across an agricultural
field before and after tillage. The after-tillage data was analyzed by
\cite{rengp} and \cite{cngp} using random sampling algorithms and a Gaussian process
of the form \eqref{eqn:gp} with
\begin{align*}
  \Ex\cpa{Z(\s)} = \beta_1\quad\text{and}\quad
  \k(d) = \exp\cpa{-\frac{d}{\l}}.
\end{align*}
We'll use the same model and data set with our deterministic algorithms. 
When we fit a sparse grid to approximate the posterior and marginalize, we get these
values for the medians
\begin{align*}
  \pa{\beta_1}_\textrm{med} = 10.86,\quad\l_\textrm{med} = 62.54,\quad\nr_\textrm{med}=0.44,\quad\textrm{and}\quad
  \var_\textrm{med} = 0.24.
\end{align*}
Figure~\ref{fig:soil-sparse-grid} plots the sparse grid constructed by Algorithm~\ref{alg:sg}; 
Figure~\ref{fig:soil-marg} plots the posterior marginalizations for $\beta_1$, $\l$, $\nr$, and 
$\var$; and Figure~\ref{fig:soil-pred} plots carbon-to-nitrogen predictions and credible sets across 
the agricultural field. [\href{https://github.com/rnburn/bbai/blob/master/example/08-soil-cn.ipynb}{source}] 

\subsection{Meuse River}\label{sec:meuse}
\begin{figure}
  \begin{subfigure}[t]{\textwidth}
    \centering
    \scalebox{1.0}{\includegraphics{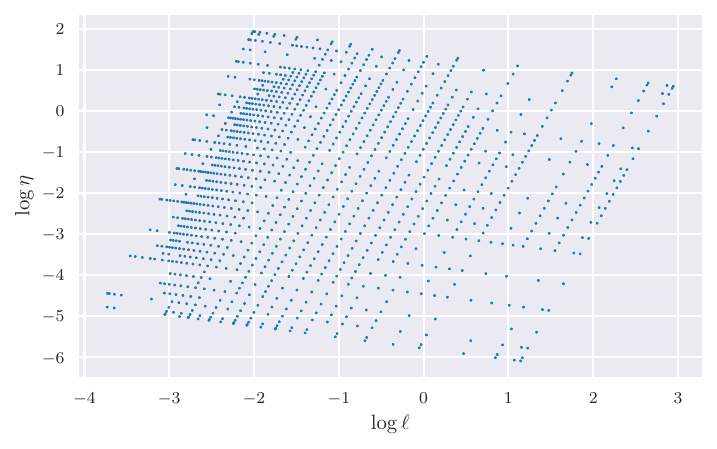}}
  \end{subfigure}
  \caption{Sparse grid used to interpolate the posterior of the Meuse data set}
  \label{fig:meuse-sparse-grid}
\end{figure}

\begin{figure}
  \centering
  \begin{subfigure}[t]{0.45\textwidth}
    \centering
    \scalebox{0.45}{\includegraphics{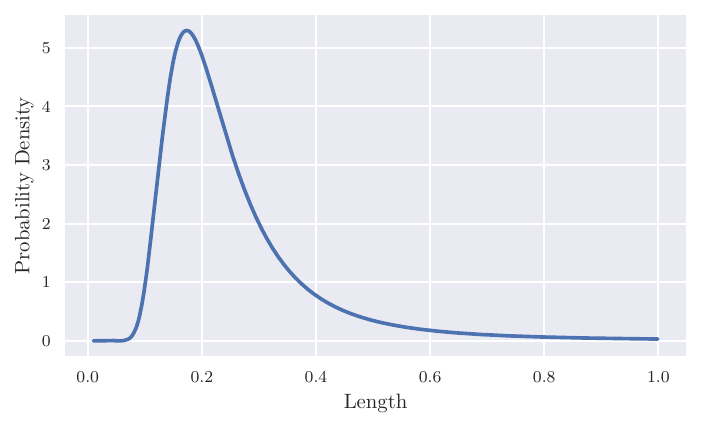}}
  \end{subfigure}%
  ~
  \begin{subfigure}[t]{0.45\textwidth}
    \centering
    \scalebox{0.45}{\includegraphics{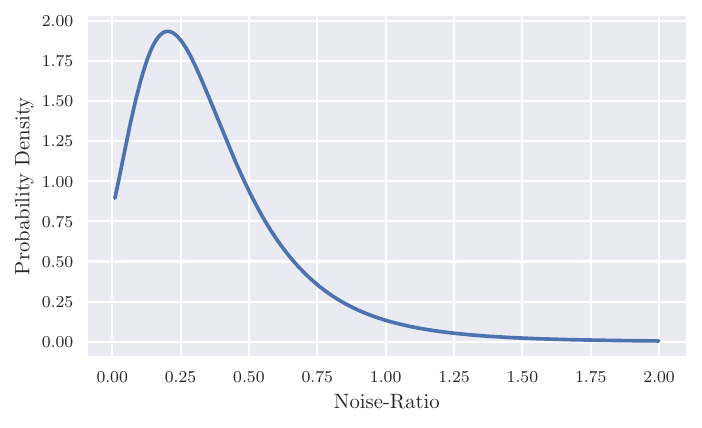}}
  \end{subfigure}
  \begin{subfigure}[t]{0.45\textwidth}
    \centering
    \scalebox{0.45}{\includegraphics{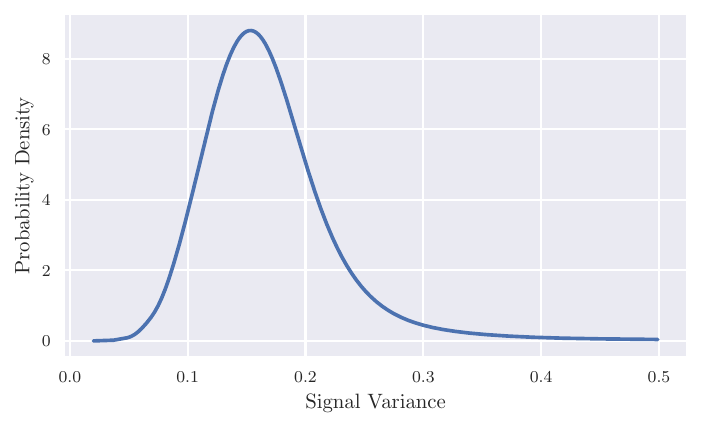}}
  \end{subfigure}%
  ~
  \begin{subfigure}[t]{0.45\textwidth}
    \centering
    \scalebox{0.45}{\includegraphics{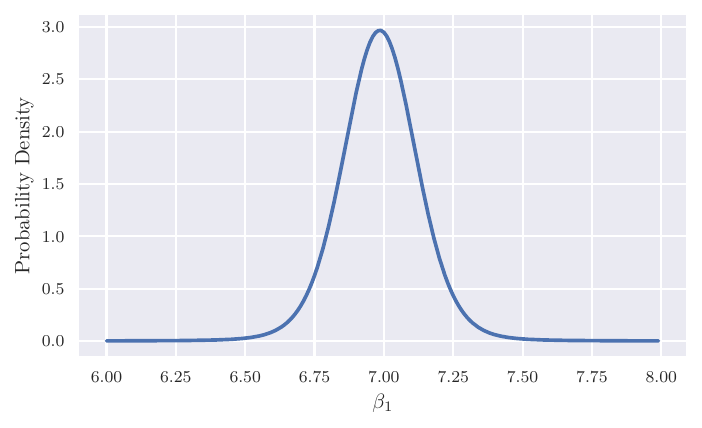}}
  \end{subfigure}
  \begin{subfigure}[t]{0.45\textwidth}
    \centering
    \scalebox{0.45}{\includegraphics{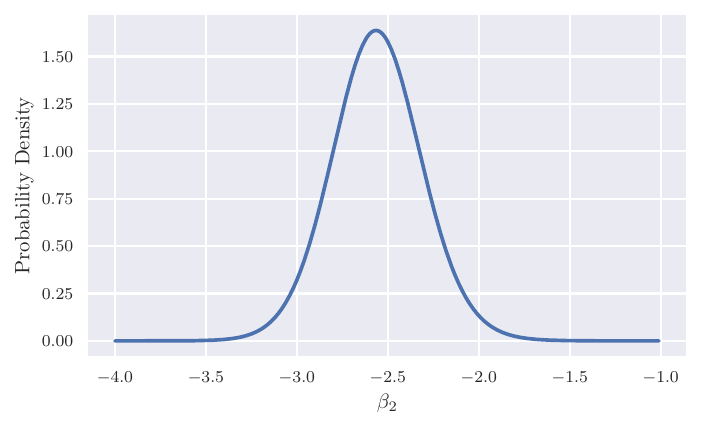}}
  \end{subfigure}
  \caption{Marginalizations of the posterior distribution of the Meuse data set}
  \label{fig:meuse-marg}
\end{figure}

\begin{figure}
  \centering
  \begin{subfigure}[t]{0.45\textwidth}
    \centering
    \scalebox{0.45}{\includegraphics{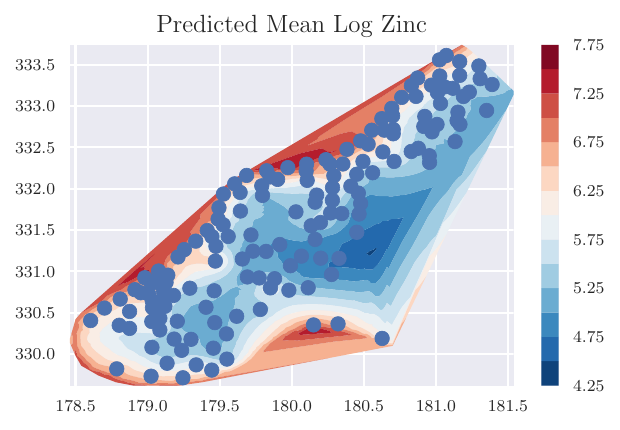}}
  \end{subfigure}%
  ~
  \begin{subfigure}[t]{0.45\textwidth}
    \centering
    \scalebox{0.45}{\includegraphics{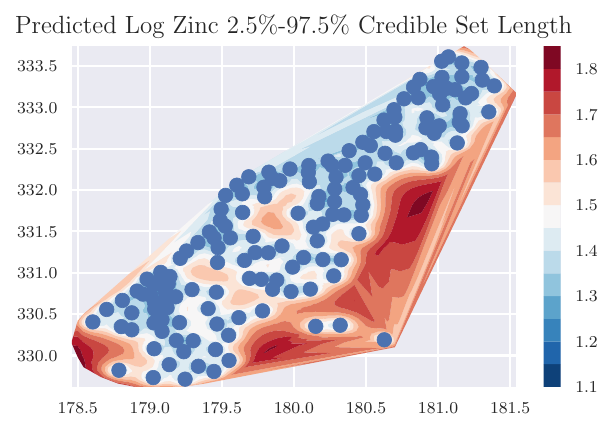}}
  \end{subfigure}
  \caption{Prediction means and credible sets of log zinc concentration for Meuse data set with sampling locations}
  \label{fig:meuse-pred}
\end{figure}

Next, we'll look at a data set from the \href{https://cran.r-project.org/web/packages/sp/index.html}{sp R-library} 
containing \num{155} measurements of zinc concentration 
(ppm) collected in a flood plain of the river Meuse \citep{rsp}. The data was previously analyzed by
\cite{meusegp} using a Gaussian process with a sampling algorithm. We'll use a similar model but
with the deterministic algorithms from \S\ref{sec:detalgo}. We model log zinc concentration as a Gaussian
process of the form \eqref{eqn:gp} with
\begin{align*}
  \Ex\cpa{Z(\s)} = \beta_1 + \beta_2 x_1(\s)\quad\text{and}\quad
  \k(d) = \exp\cpa{-\frac{d}{\l}}
\end{align*}
where $x_1(\s)$ is the square root of the distance of the flood plain sampling location, $\s$, to 
the river Meuse. After fitting the model, we compute medians
\begin{align*}
  \pa{\beta_1}_\textrm{med} = 6.99,\;\pa{\beta_2}_\textrm{med}=-2.56, \;\l_\textrm{med}=0.22,\;
  \nr_\textrm{med}=0.31,\;\textrm{and}\;\var_\textrm{med}=0.16.
\end{align*}
Figure~\ref{fig:meuse-sparse-grid} plots the sparse grid constructed by Algorithm~\ref{alg:sg}; 
Figure~\ref{fig:meuse-marg} plots the posterior marginalizations for 
$\beta_1$, $\beta_2$, $\l$, $\nr$, and $\var$; and
Figure~\ref{fig:meuse-pred} plots log zinc predictions and credible sets across the 
flood plain. 
[\href{https://github.com/rnburn/bbai/blob/master/example/11-meuse-zinc.ipynb}{source}] 

\subsection{Performance Analysis}
To get a sense of the cost of the algorithms, I measured how long it took
to apply the steps of Algorithm~\ref{alg:high} to
the soil and Meuse data sets for varying error tolerances. I computed the results 
using an 8-core AMD Ryzen 9 laptop. Table~\ref{perf:soil} and Table~\ref{perf:meuse} summarize the
performance results and provide the 25th, 50th, and 75th percentiles for the $\l$, $\nr$, and $\var$ distributions
to help show how the tolerance affects accuracy.
[\href{https://github.com/rnburn/bbai/blob/master/example/12-gaussian-process-performance.ipynb}{source}]
\begin{landscape}
\begin{table}
\centering
\begin{tabular}{lccccccccccl}\toprule
\multicolumn{3}{c}{} & 
\multicolumn{3}{c}{$\l$ Percentile} &
\multicolumn{3}{c}{$\nr$ Percentile} &
\multicolumn{3}{c}{$\var$ Percentile}
\\\cmidrule(lr){4-6}
\cmidrule(lr){7-9}
\cmidrule(lr){10-12}
$\textrm{tol}$ & \textrm{grid size} & $\textrm{elapse (s)}$ & 
25th & 50th & 75th &
25th & 50th & 75th &
25th & 50th & 75th
\\\midrule
\num{1e-2} & 249 & 1.37 
           & 42.00 & 63.50 & 106.92 
           & 0.31 & 0.45 & 0.60 
           & 0.20 & 0.25 & 0.31 \\
\num{1e-3} & 252 & 1.32 
           & 42.88 & 62.55 & 104.56 
           & 0.32 & 0.44 & 0.61 
           & 0.20 & 0.25 & 0.32 \\
\num{1e-4} & 798 & 2.86
           & 42.88 & 62.55 & 104.56 
           & 0.32 & 0.44 & 0.61 
           & 0.20 & 0.25 & 0.32 \\
\num{1e-5} & 2218 & 6.67
           & 42.88 & 62.54 & 104.57 
           & 0.32 & 0.44 & 0.61 
           & 0.20 & 0.25 & 0.32 \\
\num{1e-6} & 3415 & 9.82
           & 42.88 & 62.54 & 104.57 
           & 0.32 & 0.44 & 0.61 
           & 0.20 & 0.25 & 0.32
\\\bottomrule
\end{tabular}
\caption{Provide distribution percentiles and measure elapse time for fitting a sparse grid to the soil data set described in \S\ref{sec:soil} using various tolerances}
\label{perf:soil}
\end{table}
\begin{table}
\centering
\begin{tabular}{lccccccccccl}\toprule
\multicolumn{3}{c}{} & 
\multicolumn{3}{c}{$\l$ Percentile} &
\multicolumn{3}{c}{$\nr$ Percentile} &
\multicolumn{3}{c}{$\var$ Percentile}
\\\cmidrule(lr){4-6}
\cmidrule(lr){7-9}
\cmidrule(lr){10-12}
$\textrm{tol}$ & \textrm{grid size} & $\textrm{elapse (s)}$ & 
25th & 50th & 75th &
25th & 50th & 75th &
25th & 50th & 75th
\\\midrule
\num{1e-2} & 215 & 0.69 
           & 0.17 & 0.22 & 0.30
           & 0.17 & 0.31 & 0.50 
           & 0.13 & 0.16 & 0.20 \\
\num{1e-3} & 416 & 1.03
           & 0.17 & 0.22 & 0.30
           & 0.17 & 0.31 & 0.49
           & 0.13 & 0.16 & 0.20 \\
\num{1e-4} & 1193 & 2.42
           & 0.17 & 0.22 & 0.30
           & 0.17 & 0.31 & 0.50
           & 0.13 & 0.16 & 0.20 \\
\num{1e-5} & 2358 & 4.51
           & 0.17 & 0.22 & 0.30
           & 0.17 & 0.31 & 0.50
           & 0.13 & 0.16 & 0.20 \\
\num{1e-6} & 8666 & 17.10
           & 0.17 & 0.22 & 0.30
           & 0.17 & 0.31 & 0.50
           & 0.13 & 0.16 & 0.20
\\\bottomrule
\end{tabular}
\caption{Provide distribution percentiles and measure elapse time for fitting a sparse grid to the Meuse data set described in \S\ref{sec:meuse} using various tolerances}
\label{perf:meuse}
\end{table}
\end{landscape}

\section{Discussion}
I presented deterministic algorithms for fully Bayesian prediction and inference for spatial models. In
comparison to sampling methods such as MCMC, I would expect the algorithms to provide
more reproducible results and require less tuning, making a more turnkey approach to analysis
possible.

An area of future work could be to extend the algorithms to handle the Gaussian
process models used in model emulation and calibration. In contrast to spatial
Gaussian process models, models for emulation and calibration typically assume each
dimension of the input space has a different scale and use a product covariance function
\citep{sackscor, mengcor},
\begin{align*}
  \Cov{Z(\s), Z(\u)} &= \var \prod_{k=1}^d \psi_{\l_k}(\mid s_k - u_k\mid).
\end{align*}
\cite{sepprior} and \cite{mengsep} derived reference priors for
separable covariance functions with distinct length parameters. Provided $d$ is not too large, it may be possible
to adopt the algorithms from \S\ref{sec:detalgo} to work for this case to achieve deterministic fully 
Bayesian results. Additionally, certain modifications, such as using 
second-order information at interpolation points or assuming that the posterior is separable along the eigenvectors of its optimum's Hessian, could make the algorithms more efficient and larger values of $d$ 
possible.

\appendix
\section{Appendix: Posterior Derivatives}\label{appendix:a}
We will derive equations to compute the value, gradient, and Hessian of the negative log posterior
$\post{\l, \nr}$. From \eqref{eqn:likei} and \eqref{eqn:prior}, we have
\begin{align*}
  \post{\l, \nr} &\propto \likei{\l, \nr} \times \prior{\l, \nr} \\
                 &\propto 
                 \det{\G}^{-1/2} \det{\X' \Gi \X}^{-1/2} \pa{S^2}^{-(n - p)/2} \det{\smat}^{1/2}.
\end{align*}
Put $\hp_1 = \l$, $\hp_2 = \nr$, $\A=\X' \Gi \X$, and define
\begin{align*}
  f(\hps) = \half \log \det{\G} + \half \log \det{\A} + \divtwo{n - p} \log S^2 
          - \half\log{\det{\smat}}.
\end{align*}

Let $\LG$ denote the Cholesky factorization of $\G$, $\G = \LG\LG'$.
Then
\begin{align*}
  \A &= \X' \pa{\LG \LG'}^{-1} \X \\
     &= \X' \LG'^{-1} \LG^{-1} \X \\
     &= \pa{\LG^{-1} \X}' \pa{\LG^{-1} \X}.
\end{align*}
Let $\Q$ and $\RA$ denote the QR factorization of $\LG^{-1} \X$. Then
\begin{align*}
  \A &= \pa{\LG^{-1} \X}' \pa{\LG^{-1} \X} \\
     &= \pa{\Q \RA}' \pa{\Q \RA} \\
     &= \RA' \Q' \Q \RA \\
     &= \RA' \RA.
\end{align*}
Put $\H = \Gi \X \Ai \X' \Gi$ and $\FH = \RA'^{-1} \X' \Gi$.
Applying to $\R$ \eqref{eqn:R}, we have
$\H = \FH' \FH$ and $\R = \LG'^{-1} \LG^{-1} + \FH' \FH$.
\subsection*{Gradient}
Put
\begin{align*}
  \#1 = \half\log \det{\G},\quad
  \#2 = \half\log \det{\A},\quad
  \#3 = \divtwo{n-p} \log S^2,\ \text{and}\quad
  \#4 = \half \log\det{\smat}.
\end{align*}
Applying Jacobi's formula,
$\frac{d}{dt} \det{\bs{B}(t)} = \det{\bs{B}}\tr\cpa{\bs{B}^{-1} \frac{d \bs{B}}{dt}}$,
we have
\begin{align*}
  \pdf{\#1}{\hp_s} &= \half\tr\cpa{\G^{-1} \pdf{\G}{\hp_s}}, \\
  \pdf{\#2}{\hp_s} &= \half\tr\cpa{\A^{-1} \pdf{\A}{\hp_s}}, \\
  \pdf{\#3}{\hp_s} &= 
    \divtwo{n - p} \inv{S^2} \pdf{S^2}{\hp_s} =
    \divtwo{n - p} \inv{S^2} \y' \pdf{\R}{\hp_s} \y,\and \\
  \pdf{\#4}{\hp_s} &= \half\tr\cpa{\smat^{-1} \pdf{\smat}{\hp_s}}.
\end{align*}
Using the formula for differentiating an inverse matrix,
$\frac{d}{dt} \bs{B}(t)^{-1} = -\bs{B}^{-1} \frac{d\bs{B}}{dt} \bs{B}^{-1}$,
we derive the derivative of $\A$,
\begin{align*}
  \pdf{\A}{\phi_s} = \X' \dGis \X
\end{align*}
where
$\dGis = -\Gi \dGs \Gi$.
Differentiating $\R$ gives us
\begin{align*}
  \dRs = \pd{\hp_s} \pa{\Gi - \H} = \dGis - \dHs
\end{align*}
and
\begin{align*}
  \dHs &= \pd{\hp_s} \pa{\Gi\X\A^{-1} \X'\Gi} \\
       &=
        -\Gi \dGs \H - \H \dGs \Gi -
        \Gi \X \Ai \dAs \Ai \X\Gi \\
       &= \!\begin{multlined}[t]
        -\Gi \dGs \H - \H \dGs \Gi \\
        + \Gi \X \Ai \pa{\X' \Gi \dGs \Gi \X} \Ai \X \Gi 
      \end{multlined} \\
      &= 
        -\Gi \dGs \H - \H \dGs \Gi + \H \dGs \H \\
      &= - \pa{\Gi - \half \H} \dGs \H - \H \dGs \pa{\Gi - \half \H}.
\end{align*}
Put $\#5 = \tr\cpa{\R \dK}$.
Then
\begin{align*}
  \pa{\dsmats}_{11} &= \pd{\hp_s} \tr\cpa{\#5^2} = 2 \tr\cpa{\#5 \pdf{\#5}{\hp_s}}, \\
  \pa{\dsmats}_{12} &= \pd{\hp_s} \tr\cpa{\R^2 \dK}
                     = \tr\cpa{ \dRRs \dK + \R^2 \ddKs }, \\
  \pa{\dsmats}_{13} &=  \tr\cpa{\pdf{\#5}{\hp_s}}, \\
  \pa{\dsmats}_{22} &= \tr\cpa{\dRRs}, \\
  \pa{\dsmats}_{23} &= \tr\cpa{\dRs}, \\
  \pa{\dsmats}_{33} &= 0,
\end{align*}
and
\begin{align*}
  \pdf{\#5}{\hp_s} &= \pd{\hp_s} \pa{\R \dK} = \dRs \dK  + \R\ddKs\and \\
  \dRRs &= \dRs \R + \R \dRs.
\end{align*}

\subsection*{Hessian}
Computing second derivatives we have 
\begin{align*}
\pddf{\#1}{\hp_s}{\hp_t} &= \pd{\hp_s} \pa{\half\tr\cpa{\Gi \dGt}} \\
    &=
    \half\tr\cpa{-\Gi \dGs \Gi \dGt + \Gi \dGst}, \\
\pddf{\#2}{\hp_s}{\hp_t} &= \pd{\hp_s} \pa{\half\tr\cpa{\Ai \dAt}} \\
     &=
      \half\tr\cpa{-\Ai \dAs \Ai \dAt + \Ai \dAst}, \\
\pddf{\#3}{\hp_s}{\hp_t} &= \pd{\hp_s}\pa{ \divtwo{n - p} \inv{S^2}\dSSt} \\
    &=
    \divtwo{n - p} \pa{-\inv{S^4}\dSSs \dSSt + \inv{S^2} \dSSst},\and \\
\pddf{\#4}{\hp_s}{\hp_t} &= \pd{\hp_s} \pa{\half\tr\cpa{\smati \dsmatt}} \\
     &= \half\tr\cpa{-\smati \dsmats \smati\dsmatt + \smati \dsmatst}.
\end{align*}
For the second derivative of $\A$, we have
\begin{align*}
  \dAst &= \pd{\hp_s} \pa{-\X' \Gi \dGt \Gi \X} \\
        &= \!\begin{multlined}[t]
        \X' \Gi \dGs \Gi \dGt \Gi \X
        + \X' \Gi \dGt \Gi \dGs \Gi \X \\
        - \X' \Gi \dGst \Gi \X.
      \end{multlined}
\end{align*}
Differentiating $\R$ a second time gives us
\begin{align*}
  \dRst &= \pd{\hp_s} \pa{\dGit - \dHt} \\
        &= \dGist - \dHst
\end{align*}
where
\begin{align*}
  \dGist &= \pd{\hp_s} \pa{-\Gi \dGt \Gi} \\
         &= \!\begin{multlined}[t]\Gi \dGs \Gi \dGt \Gi +
           \Gi \dGt \Gi \dGs \Gi \\
           - \Gi \dGst \Gi\and
         \end{multlined} \\
  \dHst &= \pd{\hp_s} \pa{
  - \pa{\Gi - \half \H} \dGt \H - \H \dGt \pa{\Gi - \half \H} } \\
      &= \#D2H1 + \#D2H2 + \#D2H3
\end{align*}
and
\begin{align*}
  \#D2H1 &= -\pa{\dGis - \half \dHs} \dGt \H - \H \dGt \pa{\dGis - \half \dHs}, \\
  \#D2H2 &= -\pa{\Gi - \half \H} \dGt \dHs - \dHs \dGt \pa{\Gi - \half \H},\and \\
  \#D2H3 &= - \pa{\Gi - \half \H} \dGst \H - \H \dGst \pa{\Gi - \half \H}.
\end{align*}
Computing the second derivative of $\smat$, we have
\begin{align*}
  \pa{\dsmatst}_{11} &= 
    \pd{\hp_s} \pa{2 \tr\cpa{\#5 \pdf{\#5}{\hp_t}}} \\
      &= 2 \tr\cpa{\pdf{\#5}{\hp_s} \pdf{\#5}{\hp_t} + \#5 \pddf{\#5}{\hp_s}{\hp_t}}, \\
  \pa{\dsmatst}_{12} &= \pd{\hp_s} \pa{\tr\cpa{\dRRt \dK + \R^2 \ddKt}} \\
      &= \tr\cpa{ \dRRst \dK + \dRRs \ddKt + \dRRt \ddKs + \R^2 \ddKst}, \\
  \pa{\dsmatst}_{13} &=  \tr\cpa{\pdf{\#5}{\phi_s}{\phi_t}}, \\
  \pa{\dsmatst}_{22} &= \tr\cpa{\dRRst},\and \\
  \pa{\dsmatst}_{23} &= \tr\cpa{\dRst}
\end{align*}
and
\begin{align*}
\pddf{\#5}{\hp_s}{\hp_t} &= \pd{\hp_s} \pa{ \dRt \dK  + \R\ddKt} \\
    &= \dRst \dK  + \dRs \ddKt + \dRt\ddKs + \R\ddKst\and \\
\dRRst &= \pd{\hp_s} \pa{\dRt \R + \R \dRt} \\
       &= \pa{\dRst \R + \R \dRst} + \pa{\dRs \dRt + \dRt \dRs}.
\end{align*}

\bibliography{algo}
\end{document}